\documentclass[prd,aps,showpacs,nofootinbib,longbibliography]{revtex4}
\usepackage{amssymb,amsmath,epsfig}
\usepackage{mathrsfs}
\usepackage[usenames,dvipsnames]{color}
\usepackage[bookmarks]{hyperref}
\usepackage[svgnames]{xcolor}
\usepackage{subfigure}
\usepackage{times}
\usepackage{yhmath}
\usepackage[normalem]{ulem}

\newcommand{\Real}{\mathbb{R}}

\newcommand{\dvol}{\mbox{dvol}}

\newcommand{\proofof}[1]{\noindent {\bf Proof of #1. }}
\newcommand{\qed}{\hfill \fbox{} \vspace{.3cm}}

\newtheorem{lemma}{Lemma}

\numberwithin{equation}{section}

\begin{document}

\title{Kinetic theory for a relativistic charged gas: mathematical foundations of the hydrodynamic limit and first-order results within the projection method}

\author{Carlos Gabarrete$^1$, Ana Laura Garc\'ia-Perciante$^2$, and Olivier Sarbach$^{3}$}
\address{$^1$Departamento de Gravitaci\'on, Altas Energ\'ias y Radiaciones, Escuela de F\'isica, Facultad de Ciencias, Universidad Nacional Aut\'onoma de Honduras, Edificio E1, Ciudad Universitaria, Tegucigalpa, Francisco Moraz\'an, Honduras.}
\address{$^2$Departamento de Matem\'aticas Aplicadas y Sistemas, Universidad Aut\'onoma Metropolitana-Cuajimalpa (05348) Cuajimalpa de Morelos, Ciudad de M\'exico M\'exico.}
\address{$^3$Instituto de F\'isica y Matem\'aticas,
Universidad Michoacana de San Nicol\'as de Hidalgo,
Edificio C-3, Ciudad Universitaria, 58040 Morelia, Michoac\'an, M\'exico.}

\begin{abstract}
This work derives first-order constitutive equations for a relativistic charged gas using the Chapman–Enskog expansion of near-equilibrium solutions to the Boltzmann equation, implemented via a novel projection method. The analysis is performed in an arbitrary fixed background spacetime with an external electromagnetic field. Based on a detailed study of the linearized collision operator, we identify the trace-fixed particle frame as the most natural choice for constructing dissipative relativistic fluid theories from kinetic theory. In this frame, the state variables are defined by matching the lowest order moments of the one-particle distribution function with those of the Jüttner equilibrium distribution. The corresponding constitutive relations are obtained, and the associated transport coefficients are shown to be frame-independent when properly defined.
We also identify an additional freedom, referred to as representation freedom, and show how both the representation and frame freedoms can be implemented at the microscopic level within the projection method. This allows for a systematic derivation of general first-order constitutive equations starting from the ones obtained in the trace-fixed particle frame. For suitable parameter choices, the resulting fluid theory is strongly hyperbolic, causal, and admits stable global equilibrium states.
\end{abstract}

\date{\today}

\pacs{04.20.-q,04.40.-g, 05.20.Dd}

\maketitle
\tableofcontents

\section{Introduction}
\label{Sec:Introduction}

The kinetic theory of gases provides a formal and rigorous framework to model macroscopic phenomena based on the properties and dynamics of its individual constituents~\cite{ChapmanCowling,cercignani2012boltzmann,cercigniani1}. For dilute systems, the Boltzmann equation describes the time evolution of the one-particle distribution function when considering binary collisions, and accurately leads to the corresponding transport equations in the hydrodynamic regime when linearized around equilibrium configurations. This can be achieved by means of the perturbative approach known as the Chapman-Enskog method, which has been successfully applied to non-relativistic systems. The corresponding constitutive equations relating dissipative fluxes and forces, when truncated to first order in the gradients of the state variables, lead to the well-known Navier-Stokes equations for a compressible fluid~\cite{ChapmanCowling,cercignani2012boltzmann,cercigniani1,LaurePaper,LaureLectureNotes}. 

The traditional procedure to establish constitutive equations within the Chapman-Enskog approach has been systematized such that the existence of the solution of the linearized Boltzmann equation is enforced by eliminating the time derivatives of state variables appearing in the first-order correction to the equilibrium distribution function by means of the lower order Euler equations. This leads to the usual coupling laws (Fourier and Navier-Newton) between dissipative fluxes and spatial gradients of the state variables~\cite{ChapmanCowling,JNET24}. A more rigorous method, which we refer to as the projection method, has been discussed by Saint-Raymond in Refs.~\cite{LaurePaper,LaureLectureNotes} which, in the nonrelativistic case considered in these references, yields exactly the same constitutive equations as the ones obtained from the traditional procedure.

On the other hand, the usual linear relations predicted in the non-relativistic case, both by linear irreversible thermodynamics and kinetic theory, are inadequate in a relativistic scenario, where the second law of thermodynamics requires the dissipative fluxes to be coupled with time derivatives of the state variables, in addition to the usual spatial gradient terms. More precisely, the phenomenological approach first developed by Eckart~\cite{cE1940} leads to a heat flux driven by both the temperature gradient and the hydrodynamic acceleration. However, this proposal and its Landau-Lifshitz~\cite{LandauLifshitz-Book6} counterpart, have been shown to lead to pathological systems, due to causality issues and the presence of generic instabilities~\cite{wHlL83,wHlL85}. This problem led to the development of extended or higher-order dissipative fluid theories (see Refs.~\cite{Van2020,SZ2020,Gavassino2021,RDNR2024} for recent reviews). 

For the reasons mentioned above, first-order theories had been mostly discarded for decades. However, they have received renewed interest in recent years since it has been shown that they can lead to a physically sound system of equations without resorting to extended or higher-order formalisms~\cite{Bemfica2022,pKovtun19}. This is achieved by allowing for the appearance of non-equilibrium corrections in all the components of the particle current density and the energy-momentum-stress tensor, and for these corrections to be coupled to both first-order time and spatial derivatives of the state variables. This constitutes one of the core components of the new proposals, usually referred to as Bemfica-Disconzi-Noronha-Kovtun (BDNK) theory. The justification for the general couplings is based on two key features: i) the liberty of describing the fluid in an arbitrary frame and ii) the freedom of adding second-order on-shell corrections to the constitutive equations. Here, the choice of frame refers to the particular assignment of the state variables to non-equilibrium quantities. This choice is usually fixed through the so-called matching conditions which require certain macroscopic variables to match those corresponding to the local equilibrium configuration.\footnote{In the context of kinetic theory, this amounts to fitting certain moments of the distribution function to the ones corresponding to the local equilibrium configuration. We shall refer to these conditions in the microscopic framework as compatibility conditions.
Meanwhile, the second feature, which we refer to as the representation freedom, allows for the addition of certain combinations of time and space derivatives of the state variables which are second order when the balance equations hold.
By exploiting these features, one is led to a general first-order theory for which hyperbolicity, stability, and causality are conditioned to the fulfillment of several complicated inequalities for the parameters introduced in the proposal, and thus restricts the choice of frame and representation.}

It has been recently shown that within the \textit{trace-fixed particle frame} (TFP), only two conditions involving the parameters related to the representation choice are needed in order to achieve these desirable properties~\cite{aGjSoS2024a,aGjSoS2024b}. Furthermore, it has been shown in~\cite{aGjSoS2024a,aGjSoS2024b} that these conditions are satisfied for the particular case of a simple gas of hard spheres or disks, independently of the temperature. Such a description is obtained when the non-equilibrium variables are matched to the equilibrium configuration by fixing the particle current density and the trace of the energy-momentum-stress tensor. Although at the phenomenological level this choice of matching conditions may seem arbitrary, we here argue that it has a natural interpretation within relativistic kinetic theory.\footnote{The same matching conditions have already been introduced by John~Stewart in the context of the methods of moments, see chapter 4.4 in Ref.~\cite{Stewart-Book}.} 

In this work we perform a thorough analysis of the integral equation that is obtained when the Chapman-Enskog expansion is introduced in the relativistic Boltzmann equation. Our first important achievement consists in generalizing the projection method of Ref.~\cite{LaurePaper} to the relativistic scenario. We show that, in contrast to what occurs when the traditional Chapman-Enskog method is applied, time-derivative terms do appear in the first-order relativistic constitutive equations. Moreover, our second result consists in showing how the projection method can be performed by retaining the full frame and representation freedoms. This is achieved by exploiting the fact that the kernel of the linearized collision operator is non-trivial and by adding suitable terms to the relativistic Boltzmann equation which are identically zero when the balance equations are fulfilled. Whereas the relation between the homogeneous solution and an infinitesimal change of frame is well-known in the literature~\cite{rHpK22,Rocha2022,Rocha_2023}, we believe our method to incorporate the change of representation within kinetic theory by adding the zero terms is new. Our third important result is to prove that for the particular case of the TFP frame, the first-order Chapman-Enskog projection method leads to the fluid theory described in~\cite{aGjSoS2024a,aGjSoS2024b}, providing its microscopic foundation. Further, the possibility of deriving fluid theories in other frames is briefly discussed. Finally, our fourth result consists in proving that, including first-order off-equilibrium contributions, the Boltzmann entropy flux coincides with the Israel-Stewart entropy flux. An immediate corollary of this result is that the leading-order entropy production terms are positive-definite, as is also the case for the BDNK theory discussed in~\cite{Bemfica2022}.

It is also worthwhile to mention that a generalization of the projection method to the relativistic case was previously applied in the particular case of the Eckart frame~\cite{JNET24} in order to shed light on the possibility of writing the constitutive equations in terms of different combinations of time and spatial derivatives of the state variables within the same frame by changing the representation. Also, previous works discussing the microscopic foundation for BDNK theory within special relativity include Ref.~\cite{rHpK22} based on the Hilbert expansion and Refs.~\cite{Rocha2022,Rocha_2023} which apply other methods including a modified Chapman-Enskog procedure with a large class of compatibility conditions. Although the modified Chapman-Enskog procedure bears some similarities with our projection formalism, it is different since the latter solves the first-order equation by inverting the linearized collision operator instead of using the moments of the Boltzmann equation and is shown to generalize to arbitrary order in the Knudsen parameter.

The Chapman-Enskog projection method our article is based on relies on the observation that under suitable assumptions on the differential cross section, the linearized collision operator is self-adjoint and positive semi-definite when defined on an appropriate Hilbert space. Since its kernel is finite-dimensional, its restriction (in domain and range) to the orthogonal complement of the kernel yields a positive-definite invertible linear operator. On the one hand, this restriction guarantees the existence of the solution to the integral equation by projecting the source term onto this space. On the other hand, it also yields the uniqueness of the solution by requiring orthogonality to the kernel. As it turns out, this orthogonality condition is precisely equivalent to imposing the compatibility conditions on the first few moments of the distribution function, giving rise to the TFP frame. Since the kernel consists of the collision invariants, one can also interpret the TFP frame to be the natural generalization of the compatibility conditions in the nonrelativistic theory.


At this point it is worthwhile to explain the reason why in the Newtonian case, the projection method yields the same result as the traditional Chapman-Enskog procedure, whereas the results are different in the relativistic case: in the non-relativistic case, all the time derivatives in the source term lie in the kernel of the linearized collision operator and hence are projected to zero, leading to purely spatial gradients of the force terms. In contrast, not all the time derivative terms lie in the kernel in the relativistic theory, such that some of these terms survive the projection.

In this article, we consider a fully relativistic kinetic gas consisting of identical spinless classical point particles of mass $m$ and charge $q$  propagating in a curved spacetime background $(M,g)$ and electromagnetic field $F$. For generality, we perform the calculations in arbitrary spatial dimensions $d$. We assume that both $g$ and $F$ are smooth and that $(M,g)$ is time-oriented. We focus on the weak field hydrodynamic limit, that is, the limit in which the mean free path is much shorter than both the macroscopic scale and the characteristic length associated with the electromagnetic field. Therefore, to zeroth order, the gas lies in local thermodynamic equilibrium in which the one-particle distribution function is characterized by the particle density, the temperature, and the hydrodynamic velocity of the gas. 

The remainder of this work is organized as follows. In Section~\ref{Sec:Preliminaries} we review the relativistic Boltzmann equation and its most important properties. In Section~\ref{Sec:Equilibrium} we summarize the main features of the local equilibrium configurations, which are characterized by zero entropy production, and state some qualitative properties satisfied by the associated macroscopic observables, including their asymptotic behaviors in the low and high temperature cases. Next, we consider off-equilibrium configurations. To this purpose, in Section~\ref{Sec:HydrodynamicLimit} we discuss the weak field hydrodynamic limit of the Boltzmann equation, including the Hilbert method and the Chapman-Enskog projection method. Mathematical properties of the linearized collision operator which are fundamental for these methods to be well-defined are analyzed in Section~\ref{Sec:LinearizedCollisionTerm}. Next, in Section~\ref{Sec:ChapmanEnskog} we compute the first-order off-equilibrium contributions within the Chapman-Enskog projection method and derive the resulting dissipative fluid theory. In order to strengthen the thermodynamic interpretation of this theory, in Section~\ref{Sec:entropy} we compute the entropy production and discuss the second law of thermodynamics. Conclusions are drawn in Section~\ref{Sec:Conclusions}. Further details of our calculations and proofs of some technical lemmas in our work are found in the appendices.

Units in which the speed of light is set to one are employed, and the signature convention $(-,+,\ldots,+)$ for the metric is used. The parenthesis $(\ )$ denote the symmetric part of a tensor field. The Roman letters $\mathrm{p}$ and $\mathrm{g}$ denote the hydrostatic pressure and relative speed, respectively, whereas the Italic letters $p$ and $g$ refer to the momentum of a gas particle and the spacetime metric.

\section{Preliminaries}
\label{Sec:Preliminaries}

The state of a kinetic gas consisting of identical particles of mass $m$ and charge $q$ is described by the one-particle distribution function, that is, a function $f: \Gamma_m^+\to\Real$, $(x,p)\mapsto f(x,p)$ on the future mass shell $\Gamma_m^+$. Recall that $\Gamma_m^+$ is characterized by those pairs of points $(x,p)$ for which $x\in M$ and $p$ lies in the future mass hyperboloid $P_x^+(m)$ at $x$ (characterized by those covectors $p = p_\mu dx^\mu$ at $x$ which satisfy $p^\mu p_\mu = -m^2$ and the property that $p^\mu := g^{\mu\nu} p_\nu$ is future-directed).

The most relevant macroscopic quantities defined by $f$ are:
\begin{eqnarray}
J^\mu(f) &:=& \int\limits_{P_x^+(m)} f(x,p) p^\mu \dvol_x(p),\qquad
\hbox{(particle current density vector)},
\\
T^{\mu\nu}(f) &:=& \int\limits_{P_x^+(m)} f(x,p) p^\mu p^\nu \dvol_x(p),\qquad
\hbox{(energy-momentum-stress tensor)},\label{Eq:Energy-Momentum-StressTensor}
\\
S^\mu(f) &:=& -k_B\int\limits_{P_x^+(m)} f(x,p) \log(A f(x,p)) p^\mu  \dvol_x(p),\qquad
\hbox{(entropy current density vector)}.
\label{Eq:EntropyCurrentDensity}
\end{eqnarray}
In the expressions above, $\dvol_x(p)$ denotes the Lorentz-invariant volume element on $P_x^+(m)$. Further, $k_B$ refers to Boltzmann's constant and $A$ to an arbitrary positive constant of units $(length)^d\times (mass)^d$ that ensures that $A f$ is dimensionless. For brevity, we will often drop the explicit dependency on $f$ and just write $J^\mu$ instead of $J^\mu(f)$ and similarly for $T^{\mu\nu}$ and $S^\mu$. For a more thorough discussion on the quantities introduced here and in the rest of this section, we refer the reader to Ref.~\cite{rAcGoS2022} and references cited therein.

The dynamics of the gas is determined by the relativistic Boltzmann equation,
\begin{equation}
L_F[f] = Q[f,f],
\label{Eq:Boltzmann}
\end{equation}
where $L_F$ is the Liouville (or transport) operator, which, in terms of adapted local coordinates $(x^\mu,p_\mu)$ on the cotangent bundle, has the following representation
\begin{equation}
L_F = p^\mu\frac{\partial}{\partial x^\mu} - \frac{1}{2}\frac{\partial g^{\alpha\beta}}{\partial x^\mu} p_\alpha p_\beta\frac{\partial}{\partial p_\mu} + q F_\mu{}^\nu p_\nu\frac{\partial}{\partial p_\mu},
\end{equation}
and $Q[f,f]$ denotes the collision term, which is given by the symmetric bilinear expression
\begin{equation}
Q[f,h](x,p_1) := \frac{m^2}{2}\int\limits_{P_x^+(m)}\int\limits_{S^{d-1}}
 \mathrm{g}\sqrt{1 + \frac{\mathrm{g}^2}{4}}\frac{d\sigma}{d\Omega}(\mathrm{g},\Theta)
  \left[ f_1^* h_2^* + h_1^* f_2^* - f_1 h_2 - h_1 f_2
 \right] d\Omega(\hat{q}^*)\dvol_x(p_2).
\label{Eq:CollisionTerm}
\end{equation}
Here, $\frac{d\sigma}{d\Omega}(\mathrm{g},\Theta)$ denotes the differential cross section which is a function of the (Lorentz-invariant) relative speed $\mathrm{g} := |p_2-p_1|/m$ and the scattering angle $\Theta$. Microscopic reversibility (which is also a consequence of Lorentz-invariance) further requires $\frac{d\sigma}{d\Omega}(\mathrm{g},\Theta)$ to be invariant with respect to $\Theta \mapsto \pi - \Theta$. Further, we have abbreviated $f_j:=f(x,p_j)$, $f_j^*:=f(x,p_j^*)$, $h_j:=h(x,p_j)$, and $h_j^*:=h(x,p_j^*)$, where $(p_1,p_2)$ and $(p_1^*,p_2^*)$ denote the momenta of the particles just before and after the collision, respectively. Finally, $S^{d-1}$ denotes the $(d-1)$-dimensional unit sphere and $d\Omega(\hat{q}^*)$ the associated area element.

Multiplying both sides of Eq.~(\ref{Eq:CollisionTerm}) by a function $\Psi(x,p_1)$ on the mass shell, integrating over $P_x^+(m)$ and using the symmetries with respect to the interchange of $p_1$ with $p_2$ and/or $(p_1,p_2)$ with $(p_1^*,p_2^*)$ one obtains the identity
\begin{align}
& \int\limits_{P_x^+(m)} \Psi(x,p) Q[f,h](x,p)\dvol_x(p) \nonumber\\
 &= -\frac{1}{8}\int\limits_{P_x^+(m)}\int\limits_{P_x^+(m)}
\int\limits_{S^{d-1}}
 \mathcal{F}\frac{d\sigma}{d\Omega}\left[ \Psi_1^* + \Psi_2^* - \Psi_1 - \Psi_2 \right]
 \left[ f_1^* h_2^* + h_1^* f_2^* - f_1 h_2 - h_1 f_2 \right]
d\Omega(\hat{q}^*)\dvol_x(p_1)\dvol_x(p_2),
\label{Eq:QIntegralIdentity}
\end{align}
where for convenience we have introduced the invariant flux
\begin{equation}
\mathcal{F} := m^2\mathrm{g}\sqrt{1 + \frac{\mathrm{g}^2}{4}}
 = \sqrt{(p_1\cdot p_2)^2 - m^4}.
\label{Eq:InvariantFlux}
\end{equation}
Together with the identities (see e.g. Theorem~4 in~\cite{rAcGoS2022})
\begin{eqnarray}
\nabla^\mu J_\mu(f) &=& \int\limits_{P_x^+(m)} L_F[f] \dvol_x(p),
\label{Eq:DivIdentity1}\\
\nabla^\mu T_{\mu\nu}(f) + q F^\mu{}_\nu J_\mu(f) &=& \int\limits_{P_x^+(m)} p_\nu L_F[f] \dvol_x(p),
\label{Eq:DivIdentity2}
\end{eqnarray}
Eq.~(\ref{Eq:QIntegralIdentity}) has the following important consequences: First, considering $\Psi$ to be equal to a collision invariant $1$ or $p_\nu$, and using Eq.~(\ref{Eq:Boltzmann}), one immediately obtains that the particle current density and the energy-momentum-stress tensor associated with a solution of the relativistic Boltzmann equation satisfy the balance equations
\begin{equation}
\nabla_\mu J^\mu = 0,\qquad
\nabla_\mu T^{\mu\nu} + q  J_\mu F^{\mu\nu}= 0.
\label{Eq:Hydro}
\end{equation}
Second, choosing $\Psi = 1 + \log(A f)$ one obtains
\begin{equation}
\nabla_\mu S^\mu = \frac{k_B}{4}\int\limits_{P_x^+(m)}\int\limits_{P_x^+(m)}
\int\limits_{S^{d-1}}
 \mathcal{F}\frac{d\sigma}{d\Omega}\left[ 
 \log(A^2 f_1^* f_2^*) - \log(A^2 f_1 f_2) \right]
 \left[ f_1^* f_2^* - f_1 f_2 \right]
d\Omega(\hat{q}^*)\dvol_x(p_1)\dvol_x(p_2),
\label{Eq:HTheorem}
\end{equation}
for a solution $f$ of the relativistic Boltzmann equation. This yields the relativistic version of Boltzmann's H-theorem, namely $\nabla_\mu S^\mu\geq 0$ and $\nabla_\mu S^\mu = 0$ if and only if $\log(A f)$ is a collision invariant.\footnote{Recall that any collision invariant is a linear combination of $1$ and the components of $p$, see e.g. Theorem~5 in~\cite{rAcGoS2022}.} Therefore, the states which have zero entropy production are characterized by a local J\"uttner distribution function,
\begin{equation}
f^{(0)}(x,p) = \alpha(x) e^{\beta^\mu(x) p_\mu},
\label{Eq:f0}
\end{equation}
which is completely determined by a positive function $\alpha(x)$ and a future-directed timelike vector field $\beta^\mu(x)$ on the spacetime manifold.

\section{Local equilibrium configurations}
\label{Sec:Equilibrium}

In this section, we review some important properties of the spacetime observables associated with the local equilibrium distribution function $f^{(0)}$ defined in Eq.~(\ref{Eq:f0}). These can be computed in an elegant way by means of the generating function
\begin{equation}
Z(\alpha,\beta) := \int\limits_{P_x^+(m)} f^{(0)}(x,p) \dvol_x(p)
 = 2\alpha\left(\frac{2\pi m^2}{z} \right)^{\frac{d-1}{2}} \textbf{K}_{\frac{d-1}{2}}\left(z \right),
\label{Eq:Z0}
\end{equation}
where $z:= m\sqrt{-\beta^\mu \beta_\mu} = m/(k_B T)$ is the ratio between the particle rest mass $m$ and the thermal energy, and $\textbf{K}_\nu$ refers to the modified Bessel functions of the second kind (see~\cite{DLMF-Book} for their definition and properties). From this, one obtains the following moments of the distribution function
\begin{eqnarray}
J_\mu &=& \frac{\partial Z}{\partial\beta^\mu} = n u_\mu,
\label{Eq:T1Equilibrium}\\
T_{\mu\nu} &=& \frac{\partial^2 Z}{\partial\beta^\mu\partial\beta^\nu}
 = m n\left[ G_1 u_\mu u_\nu + z^{-1} g_{\mu\nu} \right],
\label{Eq:T2Equilibrium}\\
T_{\mu\nu\rho} &:=& \frac{\partial^3 Z}{\partial\beta^\mu\partial\beta^\nu\partial\beta^\rho}
 = m^2 n\left[ G_2 u_\mu u_\nu u_\rho + 3z^{-1} G_1 g_{(\mu\nu} u_{\rho)} \right],
\label{Eq:T3Equilibrium}\\
T_{\mu\nu\rho\sigma} &:=& \frac{\partial^4 Z}{\partial\beta^\mu\partial\beta^\nu\partial\beta^\rho\partial\beta^\sigma}
 = m^3 n\left[ G_3 u_\mu u_\nu u_\rho u_\sigma + 6 z^{-1} G_2 g_{(\mu\nu} u_\rho u_{\sigma)}  + 3z^{-2} G_1 g_{(\mu\nu} g_{\rho\sigma)} \right],
 \label{Eq:T4Equilibrium}
\end{eqnarray}
with the particle density
\begin{equation}
n := 2\alpha m\left(\frac{2\pi m^2}{z} \right)^{\frac{d-1}{2}} \textbf{K}_{\frac{d+1}{2}}(z),
\label{Eq:n0}
\end{equation}
 and where the parenthesis $(\ldots)$ denotes total symmetrization over the enclosed indices, e.g.
\begin{eqnarray}
g_{(\mu\nu} u_{\rho)} 
 &=& \frac{1}{3}\left( g_{\mu\nu} u_\rho + g_{\nu\rho} u_\mu + g_{\rho\mu} u_\nu \right),
\\
g_{(\mu\nu} u_\rho u_{\sigma)} 
 &=& \frac{1}{6}\left( 
 g_{\mu\nu} u_\rho u_\sigma + g_{\mu\rho} u_\nu u_\sigma + g_{\mu\sigma} u_\nu u_\rho 
 + g_{\nu\rho} u_\mu u_\sigma + g_{\nu\sigma} u_\mu u_\rho + g_{\rho\sigma} u_\mu u_\nu 
 \right).
\end{eqnarray}
The functions $G_k$ appearing in Eqs.~(\ref{Eq:T2Equilibrium}--\ref{Eq:T4Equilibrium}) are defined by
\begin{equation}
G_k(z) := \frac{\textbf{K}_{\frac{d+1}{2} + k}(z)}{\textbf{K}_{\frac{d+1}{2}}(z)},\qquad
k = -1,0,1,2,\ldots
\label{Eq:Gk}
\end{equation}
and they satisfy recursion relations which turn out to be useful in this work and are summarized in  Appendix~\ref{App:PropertiesGFunction}.

Equation~(\ref{Eq:T2Equilibrium}) describes the energy-momentum-stress tensor of a perfect fluid, that is $T_{\mu\nu}(f^{(0)}) = n h u_\mu u_\nu + \mathrm{p} g_{\mu\nu}$, where
\begin{eqnarray}
\label{Eq:EnthalpyPerParticle}
    h &=& e + \frac{\mathrm{p}}{n} = m G_1, \qquad \hbox{(enthalpy per particle)}, \\
\label{Eq:TotalPressure}
    \mathrm{p} &=& \frac{mn}{z} = n k_B T, \qquad \hbox{(hydrostatic pressure)}, \\
\label{Eq:Velocity}
    u^\mu &=& \frac{m}{z}\beta^\mu, \qquad \hbox{(mean particle velocity)},
\end{eqnarray}
and $e$ denotes the energy per particle. The corresponding entropy flux covector is 
\begin{equation}
S_\mu(f^{(0)})=-k_B\left[ \log(A\alpha) J_\mu + \beta^\nu T_{\mu\nu} \right],
\end{equation}
hence, $S_\mu(f^{(0)}) = s n u_\mu$ with the entropy per particle given by
\begin{equation}
s = -k_B\left[ 1 + \log(A\alpha) \right] + \frac{h}{T}.
\label{Eq:sDef}
\end{equation}
Using the relation~(\ref{Eq:n0}) between $n$, $\alpha$, and $z$ which yields
\begin{equation}
\frac{\delta n}{n} = \frac{\delta\alpha}{\alpha} + \left[ z G_1 - 1 \right]\frac{\delta T}{T},
\end{equation}
one finds that $s$ satisfies the Gibbs relation
\begin{equation}
\delta s = \frac{1}{T} \delta e + \frac{\mathrm{p}}{T} \delta\left( \frac{1}{n} \right).
\label{Eq:GibbsRelation}
\end{equation}

Instead of $\alpha$ and $\beta^\mu$, the equilibrium distribution function $f^{(0)}$ can equivalently be parametrized in terms of its particle density $n$, temperature $T$ (or the ratio between the rest mass and the thermal energy $z$), and mean particle velocity $u^\mu$. Using Eqs.~(\ref{Eq:n0}) and~(\ref{Eq:Velocity}) one finds
\begin{equation}
f^{(0)}(x,p) = \frac{n(x)}{2m (2\pi m k_B T(x))^{\frac{d-1}{2}} \textrm{\textbf{K}}_{\frac{d+1}{2}}(z)} 
\exp\left[ \frac{u^\nu(x) p_\nu}{k_B T(x)} \right].
\label{Eq:f0nTu}
\end{equation}
Alternatively, using Eq.~(\ref{Eq:sDef}) one can write
\begin{equation}
f^{(0)}(x,p) = \frac{1}{A}\exp\left[ \frac{\mu(x) + u^\nu(x) p_\nu}{k_B T(x)} - 1 \right],
\label{Eq:f0Bis}
\end{equation}
with $A$ the same constant as the one appearing in the expression for the entropy flux, Eq.~(\ref{Eq:EntropyCurrentDensity}), and $\mu := h - T s$ the Gibbs potential per particle.

For the following it will also be useful to consider the heat capacities per particle at constant pressure and volume:
\begin{eqnarray}
\label{Eq:cp}
c_{\mathrm{p}} &:=& \left. \frac{\partial h}{\partial T} \right|_{\mathrm{p}}
 = -k_B z^2 G_1' = k_B\left[ z^2 + (d+2) z G_1 - z^2 G_1^2 \right],
\\
\label{Eq:cv}
c_v &:=& \left. \frac{\partial e}{\partial T} \right|_{n} 
 = -k_B(1 + z^2 G_1') = c_{\mathrm{p}} - k_B,
\end{eqnarray}
where a prime denotes a derivative and Eq.~(\ref{Eq:G1Prime}) has been used in order to express $G'_1$ explicitly in terms of $G_1$.

Imposing the balance equations~(\ref{Eq:Hydro}) leads to the relativistic Euler equations for $n$, $T$, and $u^\mu$ which, using the notation and equations given in  Appendix~\ref{App:Macroscopic}, can be written as
\begin{eqnarray}
\dot{n} + \theta n &=& 0,
\label{Eq:Continuity0}\\
c_v\dot{T} + \theta k_B T &=& 0,
\label{Eq:Temperature0}\\
n h a^\nu + D^\nu\mathrm{p} - q n E^\nu &=& 0,
\label{Eq:Euler0}
\end{eqnarray}
where we have used Eq.~(\ref{Eq:EnthalpyPerParticle}) and the definition of $c_v$ in order to rewrite the energy conservation equation in terms of the temperature. The first two equations, together with the Gibbs relation~(\ref{Eq:GibbsRelation}), imply that $\dot{s} = 0$, such that there is no entropy production in equilibrium. However, it is important to stress that, in general, $f^{(0)}$ does not satisfy the relativistic Boltzmann equation~(\ref{Eq:Boltzmann}) unless spacetime and the electromagnetic field possess a global timelike Killing vector field (see, for instance~\cite{rAcGoS2022}).

\subsection{Nonrelativistic and ultrarelativistic limits}

In the nonrelativistic limit $z\to\infty$, and one has~\cite[Eq. 10.40.2]{DLMF}
\begin{equation}
\textbf{K}_\nu(z) = \sqrt{\frac{\pi}{2z}}e^{-z}\left[ 1 
+ \frac{4\nu^2 - 1}{8z} 
+ \frac{(4\nu^2 - 1)(4\nu^2 - 9)}{128z^2} 
+ {\cal O}\left( \frac{1}{z^3} \right) \right],
\end{equation}
which implies
\begin{equation}
G_k(z) = 1 + \frac{k(d+1+k)}{2z} 
 + \frac{k(d+1+k)}{8z^2}\left[ k(d+1+k) - 2 \right]
+ {\cal O}\left( \frac{1}{z^3} \right)
\label{Eq:GkNonrel}
\end{equation}
and
\begin{equation}
G_k'(z) = -\frac{k(d+1+k)}{2z^2} 
+ {\cal O}\left( \frac{1}{z^3} \right).
\label{Eq:GkPrimeNonrel}
\end{equation}
In particular, one finds
\begin{eqnarray}
e &=& m + \frac{d}{2} k_B T + \frac{d(d+2)}{8}\frac{(k_B T)^2}{m} + m{\cal O}\left( \frac{1}{z^3} \right),\qquad
\frac{c_v}{k_B} = \frac{d}{2} + \frac{d(d+2)}{4}\frac{k_B T}{m} + {\cal O}\left( \frac{1}{z^2} \right),
\end{eqnarray}
and $h = e + k_B T$, $c_\mathrm{p}/k_B = c_v/k_B + 1$ which, for small temperatures, reduce to the well-known Newtonian expressions.

In the ultrarelativistic limit $z\to 0$, and one has~\cite[Eq. 10.30.2]{DLMF}
\begin{equation}
z^\nu\textbf{K}_\nu(z)\to 2^{\nu-1}\Gamma(\nu),\qquad \nu > 0,
\end{equation}
which implies
\begin{equation}
\frac{G_{-1}(z)}{z}\to \frac{1}{d-1},\qquad
z^k G_k(z)\to 2^k\left( \frac{d+1}{2} \right)_k, \qquad k=0,1,2,\ldots,
\label{Eq:GkUltraRel}
\end{equation}
where $G_{-1}$ is given by Eq.~(\ref{Eq:G1G-1}). Using this and Eq.~(\ref{Eq:G1Prime}) one finds
\begin{equation}
\frac{e}{k_B T}\to d,\qquad
\frac{c_v}{k_B}\to d.
\end{equation}

\subsection{Monotonicity properties with respect to the temperature}
\label{SubSec:Monotonicity}

We close this section with a few results which will be relevant further below. They concern the monotonicity properties of the functions $G_k$ and their implications for the dependency of the quantities $h$, $e$, $c_\mathrm{p}$, $c_v$ on the temperature. The proofs, which are somehow technical, can be found in Appendix~\ref{App:GProofs}.

\begin{lemma}
\label{Lem:G-1}
Let $d\geq 2$. Then, the function $G_{-1}: [0,\infty)\to [0,1)$ is strictly monotonously increasing and invertible.
\end{lemma}

\begin{lemma}
\label{Lem:G1}
Let $d\geq 2$ and $k\in \{1,2,3,\ldots \}$. Then, the function $G_k: (0,\infty)\to (1,\infty)$ is strictly monotonously decreasing and invertible.
\end{lemma}

As a consequence of the last two lemmas on has:

\begin{lemma}
\label{Lem:hecpcv}
For fixed $n$, the generating function $Z$ defined in Eq.~(\ref{Eq:Z0}) in which $\alpha$ is eliminated using Eq.~(\ref{Eq:n0}) is monotonously decreasing in the temperature. Moreover, the functions $h,e,c_\mathrm{p},c_v$ are monotonously increasing in the temperature. Furthermore, $c_\mathrm{p}$ and $c_v$ satisfy the inequalities
\begin{equation}
\frac{d}{2} + 1 < \frac{c_\mathrm{p}}{k_B} < d + 1,\qquad
\frac{d}{2} < \frac{c_v}{k_B} < d ,\qquad T > 0,
\label{Eq:cpcvBounds}
\end{equation}
with the lower limit being attained when $T\to 0$ and the upper one when $T\to \infty$.
\end{lemma}

\section{Knudsen number, hydrodynamic limit and Chapman-Enskog approximation}
\label{Sec:HydrodynamicLimit}

From here on, we focus our attention on the hydrodynamic limit. To this purpose, in Section~\ref{SubSec:RescaledBoltzmannEq}, we first review the relevant length scales that are associated with the problem and derive the rescaled Boltzmann equation. In Section~\ref{SubSec:IIB} we discuss the weak field hydrodynamic approximation which is based on a formal expansion of the distribution function in terms of the Knudsen parameter. Next, in Section~\ref{SubSec:LinCollFormal} we introduce the linearized collision operator and discuss its most important properties. The Hilbert method, in which one obtains an infinite hierarchy of  macroscopic equations, is described in Section~\ref{SubSec:HilbertMethod}. This method is contrasted with the Chapman-Enskog one, described in Section~\ref{SubSec:CE}, which (at a given truncation order in the Knudsen parameter) leads to a closed set of macroscopic equations for the particle density, temperature, and hydrodynamic velocity. Finally, in Section~\ref{SubSec:CEVariants} we present some variants of the Chapman-Enskog method which are important for the discussion in other sections of this article.

\subsection{Rescaling of the Boltzmann equation}
\label{SubSec:RescaledBoltzmannEq}

For the following, we introduce the relevant length scales for our problem. Note that all these quantities are defined through scalars, such that they are independent of the choice of coordinates or reference frame.
\begin{itemize}
\item A characteristic length scale $\ell_\textrm{em}$ associated with the background electromagnetic field, which can be defined through the relation
\begin{equation}
\frac{q}{m}\sqrt{| F^{\mu\nu} F_{\mu\nu}|} \sim \frac{1}{\ell_{\textrm{em}}}.
\end{equation}
For example, for a purely magnetic field, $1/\ell_{\text{em}}$ is related to the cyclotron frequency (recall that we work in units for which the speed of light is one).

\item The macroscopic length scale $\ell_{\text{ms}}$, defined as a typical length over which the spacetime observables vary. For example, one could define
\begin{equation}
\ell_{\text{ms}} := \frac{n}{\sqrt{|(\nabla^\mu n)(\nabla_\mu n)|}}, \quad \hbox{with} \quad n = \sqrt{-J^\mu J_\mu},
\label{Eq:lmsdefinition}
\end{equation}
the invariant particle density.\footnote{When in local equilibrium, this $n$ coincides with the particle density $n$ defined in Eq.~(\ref{Eq:n0}).}

\item The mean free path $\ell_{\text{mfp}}$ whose nature is microscopic and describes the  average distance traveled by a particle between successive collisions. Locally, it is defined through the relation
\begin{equation}
\sigma_T \ell_{\text{mfp}} n = 1, \quad \hbox{with} \quad \sigma_T := \int\limits_{S^{d-1}} \frac{d\sigma}{d\Omega} d\Omega,
\label{Eq:ImportantRelation}
\end{equation}
the total cross section.

\item The mean inter-particle distance $\ell_\mathrm{mpd}:=n^{-1/d}$.

\item The thermal wavelength $\lambda_\mathrm{T} := h/\sqrt{2\pi m k_B T}$, where $h$ denotes Planck's constant and $T$ the temperature.

\item The curvature radius $\ell_\textrm{R}$, which can be defined through a suitable inverse power of a typical curvature invariant (see, for instance, the discussion of the geometric optics approximation in Ref.~\cite{Straumann-Book}). For example, in flat spacetime, $\ell_\textrm{R} = \infty$, whereas for a Schwarzschild black hole of mass $M$, $\ell_\textrm{R}\sim \sqrt{r^3/M}$ at a sphere of areal radius $r$.
\end{itemize}

Several of these length scales already enter when formulating the relativistic Boltzmann equation~(\ref{Eq:Boltzmann}). For instance, it is assumed that the gas particles behave as classical point particles between collisions, which requires
\begin{equation}
\lambda_\mathrm{T} \ll \ell_\mathrm{mpd}.
\end{equation}
Furthermore, one assumes that only binary elastic interactions take place and that they can be modeled as point-like collisions. This requires that the total cross section is small compared to the corresponding scale set by the curvature radius, that is,
\begin{equation}
\sigma_T\ll \ell_\mathrm{R}^{d-1}.
\end{equation}
After these observations, we may use the remaining length scales $\ell_\mathrm{ms}$, $\ell_\mathrm{em}$, and $\ell_\mathrm{mfp}$ to express Eq.~(\ref{Eq:Boltzmann}) in terms of dimensionless quantities, such that
\begin{equation}
x^\mu = \ell_{\mathrm{ms}}\overline{x}^\mu, \qquad 
p_\mu = m\overline{p}_\mu, \qquad
f = \frac{1}{\ell_{\mathrm{ms}}^d m^d} \overline{f},
\qquad
\frac{q}{m} F_{\mu\nu} = \frac{1}{\ell_{\mathrm{em}}}\overline{F}_{\mu\nu},
\qquad
\frac{d\sigma}{d\Omega} = \frac{\ell_{\mathrm{ms}}^d}{\ell_{\mathrm{mfp}}}\frac{d\overline{\sigma}}{d\Omega},
\label{Eq:DimensionlessVariables}
\end{equation}
where all the quantities with a bar are dimensionless. In this way, one obtains the rescaled Boltzmann equation:
\begin{equation}
\overline{L}[\overline{f}] + \frac{1}{\mathrm{K}_{\mathrm{em}}}\overline{V}[\overline{f}]
=
\frac{1}{\mathrm{Kn}}\overline{Q} [\overline{f},\overline{f}],
\label{Eq:DimensionlessBoltzmann}
\end{equation}
with
\begin{equation}
\overline{L} := \overline{p}^\mu\frac{\partial}{\partial \overline{x}^\mu} - \frac{1}{2}\frac{\partial g^{\alpha\beta}}{\partial \overline{x}^\mu} \overline{p}_\alpha\overline{p}_\beta\frac{\partial}{\partial\overline{p}_\mu},\qquad
\overline{V} :=  \overline{F}_\mu{}^\nu\frac{\partial}{\partial\overline{p}_\nu},
\label{Eq:LbarVbar}
\end{equation}
$\overline{Q}[\overline{f},\overline{f}]$ denoting the dimensionless collision term, and
\begin{equation}
\mathrm{Kn} := \frac{\ell_{\mathrm{mfp}}}{\ell_{\mathrm{ms}}}, \qquad
\mathrm{K}_{\mathrm{em}} := \frac{\ell_{\mathrm{em}}}{\ell_{\mathrm{ms}}}.
\end{equation}
The quantity $\mathrm{Kn}$ represents the ratio between the mean free path and the macroscopic scale and is known as the Knudsen number. Likewise, $\mathrm{K}_{\mathrm{em}}$ represents the ratio between the characteristic length scale associated with the electromagnetic field and the macroscopic scale.

Depending on whether $\mathrm{Kn}$ and $\mathrm{K}_{\mathrm{em}}$ are small or large and depending on their relative size there are different regimes one may consider. We mention the following three scenarios:
\begin{enumerate}
\item The weak field hydrodynamic approximation:
\begin{equation}
\textrm{Kn} \ll 1 \lesssim \textrm{K}_{\textrm{em}}.
\label{Eq:WeakFieldHydro}
\end{equation}
In this case the collision term dominates over the other terms. Note that these conditions imply that $\ell_{\mathrm{mfp}} \ll \ell_{\mathrm{em}}$, such that the mean free path is much smaller than the electromagnetic length scale.

\item The strong field hydrodynamic approximation:
\begin{equation}
\textrm{Kn}\sim  \textrm{K}_{\textrm{em}} \ll 1.
\label{Eq:StrongFieldHydro}
\end{equation}
In this case, both the collision term and the electromagnetic terms dominate over the transport term $\overline{L}[\overline{f}]$, and $\ell_{\text{mfp}} \sim \ell_{\mathrm{em}}$.

\item The Vlasov limit:
\begin{equation}
1\sim\textrm{K}_{\textrm{em}} \ll \textrm{Kn},
\end{equation}
in which case collisions are sub-dominant.
\end{enumerate}

In the remainder of this article, we shall focus on the first of these regimes. For approaches of regimes 2 and 3 in the nonrelativistic case, see for instance Refs.~\cite{PoP07} and \cite{Balescu-Book1}.

\subsection{The weak field hydrodynamic approximation and the Hilbert expansion}
\label{SubSec:IIB}

In the following, we study the weak field hydrodynamic regime which is characterized by Eq.~(\ref{Eq:WeakFieldHydro}) and the small parameter $\text{Kn}$. For notational simplicity we denote $\varepsilon = \text{Kn}$ and omit the bars over the rescaled quantities. The Hilbert expansion consists in seeking solutions of the rescaled Boltzmann equation~(\ref{Eq:DimensionlessBoltzmann}) in form of a power series:
\begin{equation}
f_\varepsilon = f^{(0)} + \varepsilon f^{(1)} + \varepsilon^2 f^{(2)} + \cdots,
\label{Eq:HilbertExpansion}
\end{equation}
where $f^{(0)},f^{(1)},f^{(2)},\ldots$ are determined recursively by solving the integral equations
\begin{eqnarray}
Q[f^{(0)},f^{(0)}] &=& 0,
\label{Eq:HydroLimit0}
\\
Q[f^{(0)},f^{(1)}] &=&
 \frac{1}{2} L_F[f^{(0)}],
\label{Eq:HydroLimit1}\\
Q[f^{(0)},f^{(2)}] &=&
 \frac{1}{2}L_F[f^{(1)}]
  - \frac{1}{2}Q[f^{(1)},f^{(1)}],
\label{Eq:HydroLimit2}
\\
&\ldots&
\nonumber
\end{eqnarray}
with $L_F = L + V/\textrm{K}_{\textrm{em}}$.

The zeroth-order equation~(\ref{Eq:HydroLimit0}) implies that $f^{(0)}$ must be a local J\"uttner distribution function as in Eq.~(\ref{Eq:f0}). This can be seen by applying the integral identity~(\ref{Eq:QIntegralIdentity}) to $\Psi = \log(A f)$ and $h=f=f^{(0)}$, which implies that the right-hand side of Eq.~(\ref{Eq:HTheorem}) is zero and hence that $\log(A f)$ must be a collision invariant. According to Theorem~5 in~\cite{rAcGoS2022}, this implies that $f^{(0)}$ must be precisely of the form~(\ref{Eq:f0}). At this stage, the dependency of the fields $\alpha(x)$ and $\beta^\mu(x)$ on $x$ is completely undetermined. Note that there is no entropy production at zeroth order.

\subsection{The linearized collision operator (formal definition)}
\label{SubSec:LinCollFormal}

The first-order equation~(\ref{Eq:HydroLimit1}) is an integral equation for $f^{(1)}$, and its solution requires a careful analysis of the linearized collision operator which consists of the map $f^{(1)}\mapsto Q[f^{(0)},f^{(1)}]$. In particular, the solvability of Eq.~(\ref{Eq:HydroLimit1}) requires that the source term $\frac{1}{2} L_F[f^{(0)}]$ lies in the image of this operator.

To analyze this point in more detail, it is convenient to rewrite Eq.~(\ref{Eq:HydroLimit1}) in the form
\begin{equation}
\mathcal{L}[\phi] = \mathcal{Y},
\label{Eq:FirstOrderProblem}
\end{equation}
where we have set $\phi := f^{(1)}/f^{(0)}$, $\mathcal{Y}:=-L_F[\log(A f^{(0)})]$ and the linearized collision operator $\mathcal{L}$ is formally defined as
\begin{equation}
\mathcal{L}[\phi](x,p_1)
 :=
-\frac{2Q[f^{(0)},f^{(0)}\phi](x,p_1)}{f^{(0)}(x,p_1)} 
 = \int\limits_{P_x^+(m)}\int\limits_{S^{d-1}} \mathcal{F}
 \frac{d\sigma}{d\Omega}
 f^{(0)}_2
  \left[ \phi_1 + \phi_2 - \phi_1^* - \phi_2^* \right]
d\Omega(\hat{q}^*)\dvol_x(p_2).
\label{Eq:LinCollisionOperator}
\end{equation}
Here, we have used Eq.~(\ref{Eq:CollisionTerm}) and the fact that $f^{(0)}(x,p_1^*) f^{(0)}(x,p_2^*) = f^{(0)}(x,p_1) f^{(0)}(x,p_2)$. Multiplying both sides of Eq.~(\ref{Eq:LinCollisionOperator}) by an arbitrary test function $\psi$ times $f^{(0)}$ and integrating over $p_1$ yields
\begin{eqnarray}
&& \int\limits_{P_x^+(m)} \psi(x,p)\mathcal{L}[\phi](x,p) f^{(0)}(x,p)\dvol_x(p)
\nonumber\\
&& = \frac{1}{4}\int\limits_{P_x^+(m)}\int\limits_{P_x^+(m)}\int\limits_{S^{d-1}}
 \mathcal{F}\frac{d\sigma}{d\Omega}
 f^{(0)}_1 f^{(0)}_2
  \left[ \phi_1 + \phi_2 - \phi_1^* - \phi_2^* \right]\left[ \psi_1 + \psi_2 - \psi_1^* - \psi_2^* \right]
d\Omega(\hat{q}^*)\dvol_x(p_1)\dvol_x(p_2).
\nonumber\\
\label{Eq:psiLphi}
\end{eqnarray}
This identity can also be derived directly from Eq.~(\ref{Eq:QIntegralIdentity}) by substituting $\Psi=\psi$, $f = f^{(0)}$ and $h = f^{(0)}\phi$ in this equation. Equation~(\ref{Eq:psiLphi}) shows that the linear operator $\mathcal{L}$ is symmetric with respect to the scalar product
\begin{equation}
\langle \psi, \phi \rangle_x :=
\int\limits_{P_x^+(m)} \psi(p) \phi(p) f^{(0)}(x,p) \dvol_x(p).
\label{Eq:ScalarProduct}
\end{equation}
Furthermore, for $\psi=\phi$ it follows that
\begin{equation}
\langle \phi, \mathcal{L}[\phi] \rangle_x
 = \frac{1}{4}\int\limits_{P_x^+(m)}\int\limits_{P_x^+(m)}\int\limits_{S^{d-1}}
 \mathcal{F}\frac{d\sigma}{d\Omega}
 f^{(0)}_1 f^{(0)}_2
  \left[ \phi_1 + \phi_2 - \phi_1^* - \phi_2^* \right]^2
d\Omega(\hat{q}^*)\dvol_x(p_1)\dvol_x(p_2) \geq 0,
\label{Eq:LinCollNonnegative}
\end{equation}
which shows that $\mathcal{L}$ is nonnegative and that its kernel consists of collision invariants, that is,
\begin{equation}
\ker\mathcal{L} = \mbox{span}\{ 1,p_\mu \}.
\label{Eq:LinCollKernel}
\end{equation}
Since $\mathcal{L}$ is symmetric, its image is contained in $(\ker\mathcal{L})^\perp$, the orthogonal complement of its kernel. In Section~\ref{Sec:LinearizedCollisionTerm} we will specify conditions on the differential cross section which guarantee that (when defined on a suitable function space) the image of $\mathcal{L}$ is exactly equal to $(\ker\mathcal{L})^\perp$. It then follows that the first-order problem~(\ref{Eq:FirstOrderProblem}) has a solution if and only if $\mathcal{Y}\in (\ker\mathcal{L})^\perp$, and that the solution is unique up to the addition of an element in $\ker\mathcal{L}$. Moreover, it follows that at each point $x\in M$, the operator $\mathcal{L}$ is invertible and self-adjoint when restricted to the Hilbert space $(\ker\mathcal{L})^\perp$. In the following, we shall denote by $\mathcal{L}^{-1}$ the inverse of $\mathcal{L}$ on $(\ker\mathcal{L})^\perp$.

\subsection{The Hilbert method}
\label{SubSec:HilbertMethod}

As we have just discussed, a necessary and sufficient condition for the first-order equation to be solvable is $\mathcal{Y} \in (\ker\mathcal{L})^\perp$, which is equivalent to the conditions
\begin{eqnarray}
0 &=& -\langle \mathcal{Y}, 1 \rangle_x =
\int\limits_{P_x^+(m)} L_F[f^{(0)}](x,p) \dvol_x(p) = \nabla^\mu J_\mu^{(0)}(x),
\\
0 &=& -\langle \mathcal{Y}, p_\nu \rangle_x = \int\limits_{P_x^+(m)} p_\nu L_F[f^{(0)}](x,p) \dvol_x(p) = \nabla^\mu T_{\mu\nu}^{(0)}(x) + q F^\mu{}_\nu J_\mu^{(0)}(x),
\end{eqnarray}
where we have used the identities~(\ref{Eq:DivIdentity1}) and (\ref{Eq:DivIdentity2}) for $f = f^{(0)}$ and abbreviated $J_\mu^{(0)}:=J_\mu(f^{(0)})$ and  $T_{\mu\nu}^{(0)}:=T_{\mu\nu}(f^{(0)})$. Therefore, a necessary and sufficient condition for the first-order equation to be solvable is that the macroscopic observables associated with $f^{(0)}$ satisfy the  relativistic Euler equations~(\ref{Eq:Continuity0}--\ref{Eq:Euler0}). In this sense, the Euler equations are the integrability conditions for the first-order equations.

Subject to these integrability conditions, the solution of the first-order equation~(\ref{Eq:FirstOrderProblem}) is
\begin{equation}
\phi =  \psi - \mathcal{L}^{-1}\left[ L_F(\log A f^{(0)}) \right],
\end{equation}
where $\psi\in \ker\mathcal{L}$ is a homogeneous solution of the first-order equation.

However, contrary to what occurs in the Chapman-Enskog approach discussed in the next subsection, the homogeneous solution $\psi$ is not free. Rather, it is subject to the integrability conditions of the second-order equation~(\ref{Eq:HydroLimit2}). To discuss this in a more systematic way, we set $\phi_k:=f^{(k)}/f^{(0)}$ and note that the $k$'th order equation can be written as
\begin{equation}
\mathcal{L}[\phi_k] = \mathcal{Y}_k,\qquad
\mathcal{Y}_k = \frac{1}{f^{(0)}}\left\{ -L_F[f^{(k-1)}]
 + \sum\limits_{j=1}^{k-1} Q\left[ f^{(j)},f^{(k-j)} \right]
\right\},\qquad
k = 1,2,3,\ldots
\label{Eq:HilbertkthOrder}
\end{equation}
Here, it is important to point out that the source term $\mathcal{Y}_k$ depends only on contributions $f^{(j)}$ of order $j < k$, and hence Eq.~(\ref{Eq:HilbertkthOrder}) can be solved order-by-order in $k$ provided the integrability conditions $\mathcal{Y}_k\in (\ker\mathcal{L})^\perp$ are satisfied at each order. In this case, the $k$'th order solution is
\begin{equation}
\phi_k = \psi_k + \mathcal{L}^{-1}[\mathcal{Y}_k]
\label{Eq:HilbertkthSolution}
\end{equation}
with $\psi_k\in \ker\mathcal{L}$. As we analyze in the following, the homogeneous solution $\psi_k$ is restricted by the integrability conditions for the $(k+1)$'th order equation. Taking into account that both sides of Eq.~(\ref{Eq:QIntegralIdentity}) vanish when $\Psi$ is a collision invariant, these conditions are (dropping the argument $x$ for notational simplicity)
\begin{eqnarray}
0 &=& -\langle \mathcal{Y}_{k+1}, 1 \rangle =
\int L_F[f^{(k)}](p) \dvol(p) = \nabla^\mu J_\mu^{(k)},
\label{Eq:HilbertOrthok1}\\
0 &=& -\langle \mathcal{Y}_{k+1}, p_\nu \rangle = \int p_\nu L_F[f^{(k)}](p) \dvol(p) = \nabla^\mu T_{\mu\nu}^{(k)} + q F^\mu{}_\nu J_\mu^{(k)},
\label{Eq:HilbertOrthok2}
\end{eqnarray}
where we have used once again the identities~(\ref{Eq:DivIdentity1}) and (\ref{Eq:DivIdentity2}) and abbreviated $J_\mu^{(k)} := J_\mu(f^{(k)})$ and $T_{\mu\nu}^{(k)} := T_{\mu\nu}(f^{(k)})$. To make further progress we use the definition of $\phi_k$ and Eq.~(\ref{Eq:HilbertkthSolution}) to conclude that
\begin{eqnarray}
J_\mu^{(k)} &=& \int p_\mu f^{(
0)}(p)\phi_k(x,p) \dvol(p) 
 = \langle p_\mu, \phi_k \rangle
 = \langle p_\mu, \psi_k \rangle
 + \langle p_\mu, \mathcal{L}^{-1}[\mathcal{Y}_k] \rangle,
\\
T_{\mu\nu}^{(k)} &=& \int p_\mu p_\nu f^{(0)}(p)\phi_k(x,p) \dvol(p) 
 = \langle p_\mu p_\nu, \phi_k \rangle
 = \langle p_\mu p_\nu, \psi_k \rangle
 + \langle p_\mu p_\nu, \mathcal{L}^{-1}[\mathcal{Y}_k] \rangle.
\label{Eq:Tmunuk}
\end{eqnarray}
Next, the important observation is that $\psi_k\in \ker\mathcal{L}$ if and only if
\begin{equation}
\psi_k(p) = a_k + b_k^\mu p_\mu
\end{equation}
for some function $a_k$ and a vector field $b_k^\mu$ on the spacetime manifold (see Eq.~(\ref{Eq:LinCollKernel})), and hence
\begin{equation}
\psi_k(p) f^{(0)}(p) = (a_k + b_k^\mu p_\mu) \alpha e^{\beta^\mu p_\mu}
 = \left. \frac{d}{d\lambda} \alpha_k(\lambda) e^{\beta_k(\lambda)^\mu p_\mu} \right|_{\lambda=0},
\label{Eq:MaxwellJuttnerVariation}
\end{equation}
with $\alpha_k(\lambda) = \alpha(1 + \lambda a_k)$ and $\beta_k(\lambda)^\mu = \beta^\mu + \lambda b_k^\mu$. It follows from these considerations that $\psi_k\in \ker\mathcal{L}$ if and only if $\psi_k f^{(0)}$ is the first variation of $f^{(0)}$ within the class of local J\"uttner distribution functions. Therefore, the contribution arising from the homogeneous solution $\psi_k$ leads to an infinitesimal change of frame~\cite{rHpK22}. Taking this into account, one can write
\begin{equation}
\langle p_\mu,\psi_k \rangle = \left. \frac{d}{d\lambda} \tilde{J}_\mu^{(k)}(\lambda) \right|_{\lambda=0},
\qquad
\langle p_\mu p_\nu,\psi_k \rangle = \left. \frac{d}{d\lambda} \tilde{T}_{\mu\nu}^{(k)}(\lambda) \right|_{\lambda=0},
\end{equation}
where $\tilde{J}_\mu^{(k)}(\lambda)$ and $\tilde{T}_{\mu\nu}^{(k)}(\lambda)$ refer to the current density and energy-momentum-stress tensor associated with the distribution function $\alpha_k(\lambda) e^{\beta_k(\lambda)^\mu p_\mu}$. Hence, it follows from Eqs.~(\ref{Eq:HilbertOrthok1}--\ref{Eq:Tmunuk}) that the integrability conditions for the $(k+1)$'th order equation can be written as
\begin{eqnarray}
\left. \frac{d}{d\lambda} \nabla^\mu \tilde{J}_\mu^{(k)}(\lambda) \right|_{\lambda=0} &=& \mathcal{K},
\label{Eq:LinContinuity}\\
\left. \frac{d}{d\lambda}\left[ \nabla^\mu \tilde{T}_{\mu\nu}^{(k)}(\lambda) + q F^\mu{}_\nu \tilde{J}_\mu^{(k)}(\lambda) \right] \right|_{\lambda=0} &=& \mathcal{K}_\nu,
\label{Eq:LinEuler}
\end{eqnarray}
with the source terms
\begin{eqnarray}
\mathcal{K} &:=& -\nabla^\mu \langle p_\mu, \mathcal{L}^{-1}[\mathcal{Y}_k] \rangle,\\
\mathcal{K}_\nu &:=& -\nabla^\mu \langle p_\mu p_\nu, \mathcal{L}^{-1}[\mathcal{Y}_k] \rangle - q F^\mu{}_\nu\langle p_\mu, \mathcal{L}^{-1}[\mathcal{Y}_k] \rangle,
\end{eqnarray}
which only depend on contributions $f^{(j)}$ with $j < k$. Equations~(\ref{Eq:LinContinuity}) and (\ref{Eq:LinEuler}) are the linearized Euler equations for a perfect fluid in the presence of the source terms $\mathcal{K}$ and $\mathcal{K}_\nu$.

Summarizing, the Hilbert expansion leads to the (perfect fluid) relativistic Euler equations for $f^{(0)}$ and to an infinite family of linearized Euler equations for $\psi_k f^{(0)}$ in the presence of source terms, which can be solved order-by-order in $k$. At each order the solution is given by Eq.~(\ref{Eq:HilbertkthSolution}) and the corresponding contributions to the observables $J_\mu^{(k)}$ and $T_{\mu\nu}^{(k)}$ satisfy the homogeneous Euler equations. The homogeneous part $\psi_k$ describes an infinitesimal change of frame and is determined by the integrability conditions for the next-order equation.

For rigorous results on the hydrodynamic limit of the relativistic Boltzmann equation which are based on the Hilbert method, see for instance Refs.~\cite{jSrS11,yGqX21}. For other discussions regarding the relativistic Hilbert method and its limitations, see Refs.~\cite{CercignaniKremer-Book} and \cite{rHpK22}.

\subsection{The Chapman-Enskog projection method}
\label{SubSec:CE}

The Chapman-Enskog approach is slightly different from the Hilbert method discussed in the previous section. Although it is based on an expansion of the one-particle distribution function similar to  Eq.~(\ref{Eq:HilbertExpansion}), the crucial difference relies on the fact that the function $\alpha(x)$ and the timelike vector field $\beta^\mu(x)$ parametrizing $f^{(0)}$ through Eq.~(\ref{Eq:f0}) are determined by the (yet unknown) full distribution function $f$ through suitable compatibility conditions, instead of the Euler equations. The methodology we describe next follows the formal approach described in Ref.~\cite{LaurePaper}, here generalized to the relativistic regime and considering a charged fluid in the presence of a background electromagnetic field.

Let $f_\varepsilon$ be a one-parameter family of solutions of the rescaled Boltzmann equations
\begin{equation}
L_F[f_\varepsilon] = \frac{1}{\varepsilon} Q[f_\varepsilon,f_\varepsilon].
\label{Eq:BoltzmannBis}
\end{equation}
The Chapman-Enskog ansatz is given by~\cite{LaurePaper,LaureLectureNotes}
\begin{equation}
f_\varepsilon = f^{(0)}_\varepsilon + \varepsilon h_\varepsilon,
\label{Eq:CEAnsatz}
\end{equation}
where $f^{(0)}_\varepsilon$ denotes a local J\"uttner distribution function of the form of Eq.~(\ref{Eq:f0nTu}) and $h_\varepsilon$ determines the off-equilibrium correction. Here, the subscript $\varepsilon$ in $f^{(0)}_\varepsilon$ indicates that the local equilibrium function $f^{(0)}_\varepsilon$ depends on $\varepsilon$ through the fields $n_\varepsilon(x)$, $T_\varepsilon(x)$, and $u^\mu_\varepsilon(x)$ that parametrize it by means of suitable compatibility conditions which specify a unique frame. Hence, in contrast to the Hilbert method in which the local equilibrium contribution $f^{(0)}$ is completely determined by the ideal Euler equations, in the Chapman-Enskog method the fields $n_\varepsilon(x)$, $T_\varepsilon(x)$, and $u^\mu_\varepsilon(x)$ that parametrize the local distribution function $f_\varepsilon^{(0)}$ cannot be obtained from the zeroth-order equations alone. Instead, they are determined from the balance equations at the end of the calculation, once $f_\varepsilon$ (or one if its truncations) is known.

Introducing the Chapman-Enskog ansatz~(\ref{Eq:CEAnsatz}) into Eq.~(\ref{Eq:BoltzmannBis}) and recalling the definition~(\ref{Eq:LinCollisionOperator}) of the linearized collision operator yields
\begin{equation}
\mathcal{L}_\varepsilon\left[ \frac{h_\varepsilon}{f_\varepsilon^{(0)}} \right] = -L_F[\log(Af_\varepsilon^{(0)})]
 - \varepsilon\frac{L_F[h_\varepsilon] - Q[h_\varepsilon,h_\varepsilon]}{f^{(0)}_\varepsilon},
\label{Eq:CEImplicit}
\end{equation}
where $\mathcal{L}_\varepsilon$ refers to the linearized collision operator at $f_\varepsilon^{(0)}$. As discussed previously, the integrability condition for this equation is that the right-hand side lies orthogonal to $\ker\mathcal{L}_\varepsilon$. In the Chapman-Enskog approach as described in~\cite{LaurePaper} this is enforced by taking the orthogonal projection of the right-hand side of Eq.~(\ref{Eq:CEImplicit}) onto $(\ker\mathcal{L}_\varepsilon)^\perp$. Since $Q[h_\varepsilon,h_\varepsilon]/f_\varepsilon^{(0)}$ is already orthogonal to $\ker\mathcal{L}_\varepsilon$ (see Eq.~(\ref{Eq:QIntegralIdentity})), this projection is valid as long as
\begin{equation}
0 = \left\langle \Psi, L_F[\log(Af_\varepsilon^{(0)}] + \varepsilon\frac{L_F[h_\varepsilon]}{f_\varepsilon^{(0)}} \right\rangle
 = \left\langle \Psi, \frac{L_F[f_\varepsilon]}{f_\varepsilon^{(0)}} \right\rangle
 = \int \Psi(p) L_F[f_\varepsilon](p)\dvol(p),
\end{equation}
for all collision invariants $\Psi$, which is equivalent to the statement that $J_\varepsilon^\mu := J^\mu(f_\varepsilon)$ and $T_\varepsilon^{\mu\nu} := T^{\mu\nu}(f_\varepsilon)$ satisfy the balance equations. Therefore, provided one ensures that the observables associated with $f_\varepsilon$ satisfy the balance equations~(\ref{Eq:Hydro}) at the end of the process, one is allowed to take the orthogonal projection of the right-hand side of (\ref{Eq:CEImplicit}) on $(\ker\mathcal{L}_\varepsilon)^\perp$, such that
\begin{equation}
\mathcal{L}_\varepsilon\left[ \frac{h_\varepsilon}{f_\varepsilon^{(0)}} \right] = -\left( L_F[\log(Af_\varepsilon^{(0)}] \right)^\perp
 - \varepsilon\left( \frac{L_F[h_\varepsilon] - Q[h_\varepsilon,h_\varepsilon]}{f^{(0)}_\varepsilon} \right)^\perp,
\label{Eq:CEImplicitProjected}
\end{equation}
where $(\ldots)^\perp$ denotes the orthogonal projection of $(\ldots)$ onto $(\ker\mathcal{L}_\varepsilon)^\perp$.

As discussed at the end of Section~\ref{SubSec:LinCollFormal}, when restricted to the Hilbert space $(\ker\mathcal{L}_\varepsilon)^\perp$, the operator $\mathcal{L}_\varepsilon$ is self-adjoint and invertible, and we denote the inverse of this restriction by $\mathcal{L}_\varepsilon^{-1}$. Therefore, we obtain from Eq.~(\ref{Eq:CEImplicitProjected}) the following implicit equation for $h_\varepsilon$
\begin{equation}
h_\varepsilon = h_0 - \varepsilon f^{(0)}_\varepsilon \mathcal{L}_\varepsilon^{-1}\left( \frac{L_F[h_\varepsilon] - Q[h_\varepsilon,h_\varepsilon]}{f^{(0)}_\varepsilon} \right)^\perp,
\label{Eq:CEImpplicitInverted}
\end{equation}
where
\begin{equation}
h_0 := -f^{(0)}_\varepsilon \mathcal{L}_\varepsilon^{-1}\left[ L_F\log(A f^{(0)}_\varepsilon) \right]^\perp
\label{h0}
\end{equation}
represents the leading-order off equilibrium correction term. At this point, it is important to realize that by passing from Eq.~(\ref{Eq:CEImplicitProjected}) to Eq.~(\ref{Eq:CEImpplicitInverted}) we have made a (very natural!) choice of frame. Indeed, since by definition the range of $\mathcal{L}_\varepsilon^{-1}$ lies in $(\ker\mathcal{L}_\varepsilon)^\perp$, it follows that $h_\varepsilon/f_\varepsilon^{(0)}$ also lies in this space. In view of Eq.~(\ref{Eq:CEAnsatz}) this means that we have imposed the compatibility conditions:
\begin{equation}
\int h_\varepsilon \dvol(p) = 0,
\qquad
\int p_\mu h_\varepsilon \dvol(p) = 0,\label{Eq:TFPF0}
\end{equation}
or
\begin{eqnarray}
&& \int f_\varepsilon \dvol(p) = \int f_\varepsilon^{(0)} \dvol(p),
\label{Eq:TFPF1}\\
&& \int p_\mu f_\varepsilon \dvol(p) = \int p_\mu f_\varepsilon^{(0)} \dvol(p).
\label{Eq:TFPF2}
\end{eqnarray} 

In principle, one can try to solve Eq.~(\ref{Eq:CEImpplicitInverted}) by expanding
\begin{equation}
h_\varepsilon = \sum\limits_{n=0}^\infty \varepsilon^n h_n,
\end{equation}
which leads to the iteration
\begin{equation}
h_{n+1} = -f^{(0)}_\varepsilon \mathcal{L}_\varepsilon^{-1}\left[ \frac{L_F[h_n] - \sum\limits_{k=0}^n Q[h_k,h_{n-k}]}{f^{(0)}_\varepsilon} \right]^\perp,\qquad
n=0,1,2,\ldots
\end{equation}
and which can be formally solved order-by-order. However, to our knowledge, there are no results on the convergence of the resulting series, so it would be interesting to analyze this problem. Note that by construction, $h_n$ depends on the spacetime derivatives of $n_\varepsilon$, $T_\varepsilon$, and $u_\varepsilon^\mu$ up to order $n+1$.

Once the correction term $h_\varepsilon$ has been determined from $f_\varepsilon^{(0)}$ (up to certain order in $\varepsilon$, when truncated) one obtains $n_\varepsilon$, $T_\varepsilon$, and $u_\varepsilon^\mu$ from the balance equations~(\ref{Eq:Hydro}). Since the compatibility conditions~(\ref{Eq:TFPF2}) imply $J^\mu(h_\varepsilon) = 0$, these equations read
\begin{eqnarray}
&& \nabla_\mu J^\mu(f^{(0)}_\varepsilon) = 0,
\label{Eq:ContinuityEps}\\
&& \nabla_\mu T^{\mu\nu}(f^{(0)}_\varepsilon) + q F_\mu{}^\nu J^\mu(f^{(0)}_\varepsilon) 
 = -\varepsilon\nabla_\mu T^{\mu\nu}(h_\varepsilon)
 = -\sum\limits_{n=0}^\infty \varepsilon^{n+1} \nabla_\mu T^{\mu\nu}(h_n),
\label{Eq:EulerEps}
\end{eqnarray}
where the right-hand side of Eq.~(\ref{Eq:EulerEps}) describes the dissipative terms. Note that these non-equilibrium contributions depend on derivatives of $n_\varepsilon$, $T_\varepsilon$, and $u_\varepsilon^\mu$ of the order $\geq 2$, whereas the left-hand sides of Eqs.~(\ref{Eq:ContinuityEps}) and (\ref{Eq:EulerEps}) only depend on first-derivatives of these fields. In this sense, the left-hand side is of the order of the wave number $k$, whereas the right-hand side is of the order of $\varepsilon k^2 + \mathcal{O}(\varepsilon^2 k^3)$. When written in physical units, $\varepsilon k = \ell_{\mathrm{mfp}} k^{(phys)}$ and hence the right-hand side is a small correction when $k^{(phys)}\ll 1/\ell_{\mathrm{mfp}}$. Using the notation and balance equations summarized in Appendix~\ref{App:Macroscopic}, the first-order equations yield the following system for the state variables $(n,T,u^\mu)$:
\begin{eqnarray}
\dot{n} + \theta n &=& 0,
\label{Eq:ContinuityDiss}\\
\dot{T} + \frac{k_B}{c_v}\theta T  
 &=& -\frac{1}{n c_v}\left[ 
 \dot{\epsilon}^{(1)} + \frac{d+1}{d}\theta\epsilon^{(1)}
 + (D^\mu + 2a^\mu)\mathcal{Q}^{(1)}_\mu + \sigma^{\mu\nu}\mathcal{T}^{(1)}_{\mu\nu} \right],
\label{Eq:HeatDiss}\\
a_\nu + \frac{D_\nu \mathrm{p}}{nh} - \frac{q}{h} E_\nu &=& -\frac{1}{nh}\left[ \frac{1}{d} D_\nu\epsilon^{(1)} + \frac{d+1}{d} a_\nu\epsilon^{(1)} + \dot{\mathcal{Q}}^{(1)}_\nu + \frac{d+1}{d}\theta\mathcal{Q}^{(1)}_\nu + (\sigma^{\mu\nu} + \omega^{\mu\nu})\mathcal{Q}^{(1)}_\mu + (D^\mu + a^\mu)\mathcal{T}^{(1)}_{\mu\nu}
\right],\qquad\quad
\label{Eq:EulerDiss}
\end{eqnarray}
where we have used the fact that in the TFP frame $\mathcal{N} = n$, $\mathcal{J}^\mu = 0$, $\mathcal{E} = n e +  \epsilon^{(1)}$, and $\mathcal{P} = \mathrm{p} + \epsilon^{(1)}/d$ with $\epsilon^{(1)}$ denoting the off-equilibrium contribution to the energy density, such that
\begin{equation}
\varepsilon T_{\mu\nu}(h_0) = \epsilon^{(1)} \left( u_\mu u_\mu + \frac{1}{d}\Delta_{\mu\nu} \right) + 2 u_{(\mu} \mathcal{Q}^{(1)}_{\nu)} + \mathcal{T}^{(1)}_{\mu\nu}.
\label{Eq:epsilonTmunu}
\end{equation}
Here and in the rest of this work, $\Delta_{\mu\nu} := g_{\mu\nu} + u_\mu u_\nu$.

In the following, we shall restrict ourselves to the first-order truncation consisting of Eqs.~(\ref{Eq:ContinuityDiss}--\ref{Eq:EulerDiss}) together with the constitutive equations which relate $\epsilon^{(1)}$, $\mathcal{Q}^{(1)}_{\nu}$, and $\mathcal{T}^{(1)}_{\mu\nu}$ with the gradients of $n$, $T$, and $u^\mu$ according to Eq.~(\ref{Eq:epsilonTmunu}) with $h_0$ given by Eq.~(\ref{h0}).

\subsection{Changes of frame and representation}
\label{SubSec:CEVariants}

We conclude this section by mentioning two possible types of generalizations of the Chapman-Enskog method. As discussed later, these generalizations are important for the frame-invariant interpretation of first-order dissipative fluids which have attracted much attention in recent years.

The first type of generalization is related to the choice of frame. Recall that when passing from Eq.~(\ref{Eq:CEImplicitProjected}) to Eq.~(\ref{Eq:CEImpplicitInverted}) we have completely fixed the frame by requiring that $h_\varepsilon/f_\varepsilon^{(0)}$ is orthogonal to the kernel of the linearized collision operator $\mathcal{L}_\varepsilon$ which led to the compatibility conditions~(\ref{Eq:TFPF1}) and (\ref{Eq:TFPF2}). However, one can add to the right-hand side of Eq.~(\ref{Eq:CEImpplicitInverted}) any term of the form
\begin{equation}
f_\varepsilon^{(0)}\psi_\varepsilon,\qquad
\psi_\varepsilon\in \ker\mathcal{L}_\varepsilon,
\end{equation}
since with the addition of this term Eq.~(\ref{Eq:CEImpplicitInverted}) still implies Eq.~(\ref{Eq:CEImplicitProjected}). As discussed below Eq.~(\ref{Eq:MaxwellJuttnerVariation}) such a term corresponds to a variation of the equilibrium configuration, and hence it represents a change of frame. Therefore, by choosing $\psi_\varepsilon$ appropriately, one should in principle be able to derive the results in any desired frame. This will be exemplified in the context of the first-order theory in Section~\ref{Sec:Invariant}.

The second generalization of the Chapman-Enskog method adds the following terms to the right-hand side of Eq.~(\ref{Eq:CEImplicitProjected}):
\begin{equation}
-\frac{1}{k_B T_\varepsilon n_\varepsilon}\left[ \hat{\Gamma}_0(\nabla_\mu J_\varepsilon^\mu)u_\alpha 
 + \frac{1}{k_B T_\varepsilon}\left( \nabla_\mu T_\varepsilon^{\mu\nu} + q F_\mu{}^\nu J_\varepsilon^\mu \right) \left( \hat{\Gamma}_1 u_\nu u_\alpha  + \hat{\Gamma}_2\Delta_{\nu\alpha} \right) \right] u_\beta(p^\alpha p^\beta)^\perp,
\label{Eq:repchange}
\end{equation}
where $\hat{\Gamma}_0$, $\hat{\Gamma}_1$, and $\hat{\Gamma}_2$ are dimensionless free parameters that can, in principle, depend on the temperature parameter $T_\varepsilon$. Since $J_\varepsilon^\mu$ and $T_\varepsilon^{\mu\nu}$ are required to solve the balance equations, these terms are exactly zero {\em a posteriori}. Furthermore, since by construction they are orthogonal to $\ker\mathcal{L}_\varepsilon$, the integrability condition for passing from Eq.~(\ref{Eq:CEImplicitProjected}) to (\ref{Eq:CEImpplicitInverted}) is still satisfied. As discussed later, the addition of these terms corresponds to a change of representation when truncating the series to first order, which turns out to be crucial to obtain hyperbolic evolution equations.

\section{Mathematical discussion of the linearized collision operator}
\label{Sec:LinearizedCollisionTerm}

The considerations in Section~\ref{SubSec:LinCollFormal} motivate a well-defined mathematical framework that allows one to establish rigorous results regarding the properties of the linearized collision operator $\mathcal{L}$. A detailed discussion of such results lies beyond the scope of this article and have already been analyzed elsewhere, see for instance~\cite{mDmE88} for the three-dimensional case and \cite{LaureLectureNotes} and references therein for the non-relativistic limit. However, since the properties of $\mathcal{L}$ play such a central role in our analysis, we wish to provide a brief sketch of the methods used and state the most important results that are known so far.

For the following discussion we fix an event $x\in M$ and an equilibrium distribution function $f^{(0)}$ of the J\"uttner form~(\ref{Eq:f0}) in which it is assumed that $\alpha(x)\neq 0$ and $\beta^\mu(x) = u^\mu(x)/(k_B T(x))$ is future-directed timelike with $u^\mu$ normalized such that $u^\mu u_\mu = -1$. Next, we introduce the separable  Hilbert space\footnote{Note that if $\psi$ is represented in terms of an orthonormal frame at $x$ whose timelike vector coincides with $u^\mu(x)$, then $\mathcal{H}_x$ is the weighted $L^2$-space
$$
\left\{ \psi: \Real^d\to \Real : \int\limits_{\Real^d} |\psi(p)|^2 e^{-\frac{\sqrt{|p|^2 + m^2}}{k_B T(x)}} \frac{d^d p}{\sqrt{|p|^2 + m^2}} < \infty \right\}.
$$
} ${\cal H}_x := L^2(P_x^+(m), d\mu_x)$ consisting of measurable functions $\psi: P_x^+(m)\to \Real$ on the future mass hyperboloid $P_x^+(m)$ which are Lebesgue-square integrable with respect to the measure $d\mu_x(p) := f^{(0)}(x,p)\dvol_x(p)$. The scalar product on ${\cal H}_x$ is defined by Eq.~(\ref{Eq:ScalarProduct}), that is,
\begin{equation}
\langle \psi, \phi \rangle_x :=
\int\limits_{P_x^+(m)} \psi(p) \phi(p) d\mu_x(p),\qquad
\phi,\psi\in {\cal H}_x.
\label{Eq:ScalarProductBis}
\end{equation}
The goal is to prove that the linearized collision operator $\mathcal{L}$ which is formally given by Eq.~(\ref{Eq:LinCollisionOperator}) defines a nonnegative self-adjoint operator $\mathcal{L}: D(\mathcal{L})\subset {\cal H}_x\to {\cal H}_x$, with domain $D(\mathcal{L})$, which possesses a spectral gap (that is, a smallest positive element $\nu_0 > 0$ of the spectrum). It then follows from the spectral theory of self-adjoint operators~\cite{ReedSimon80I,ReedSimon80IV} that the restriction of $\mathcal{L}$ to the Hilbert space $(\ker\mathcal{L})^\perp\subset \mathcal{H}_x$ is again self-adjoint and has positive spectrum contained in $[\nu_0,\infty)$. In particular, this restriction has a well-defined inverse
\begin{equation}
\mathcal{L}^{-1}: (\ker\mathcal{L})^\perp\to (\ker\mathcal{L})^\perp\cap D(\mathcal{L}),
\end{equation}
which was already denoted by $\mathcal{L}^{-1}$ in the previous section. For the purpose of this article this is the most important result of this section, since it provides the basis for a rigorous treatment of the Chapman-Enskog and Hilbert methods discussed in the previous section.

In order to prove the above properties and determine $D(\mathcal{L})$ one needs to make suitable assumptions on the differential cross section. To keep the discussion simple, we do not specify the most general possible conditions and restrict ourselves to the following assumptions: there are constants $\sigma_1 > 0$, $\sigma_2 > 0$, and $0\leq\rho < 2$, such that for all $\mathrm{g}\geq 0$
\begin{equation}
\sigma_T(\mathrm{g}) := \int\limits_{S^{d-1}} \frac{d\sigma}{d\Omega}(\mathrm{g},\Theta) d\Omega \geq \sigma_1\frac{\mathrm{g}^{\rho+1}}{1+\mathrm{g}},\qquad
\frac{d\sigma}{d\Omega}(\mathrm{g},\Theta) \leq \frac{\sigma_2}{\mbox{Vol}(S^{d-1})}\frac{1 + \mathrm{g}^{\rho+1}}{\mathrm{g}}.
\label{Eq:GradCutOff1}
\end{equation}
Here, the first inequality  is a consequence of the ``hard interaction" assumption~(4.7) made in~\cite{mDmE88}. On the other hand, in the three-dimensional case the second inequality is equivalent to the assumption~(2.7) in~\cite{mDmE88} with $\alpha=1$ and $\gamma=0$, and it describes a relativistic generalization of Grad's cutoff process which tames the singularities in the cross section originating from grazing collisions. Since only the combination $\mathcal{F} \frac{d\sigma}{d\Omega}$ appears in the linearized collision operator it is convenient to multiply both sides of the inequalities in Eq.~(\ref{Eq:GradCutOff1}) by the invariant flux $\mathcal{F} = m^2\mathrm{g}\sqrt{1 + \mathrm{g}^2/4}$, which leads to the equivalent conditions
\begin{equation}
\mathcal{F}\sigma_T(\mathrm{g}) \geq c_1\mathrm{g}^{\rho+2},
\qquad
\mathcal{F}\frac{d\sigma}{d\Omega}(\mathrm{g},\Theta) \leq \frac{c_2}{\mbox{Vol}(S^{d-1})}\left( 1 + \mathrm{g}^{\rho+2} \right),
\label{Eq:GradCutOff2}
\end{equation}
for constants $c_1 > 0$, $c_2 > 0$, and $0\leq\rho < 2$. Integrating the second inequality over $S^{d-1}$ implies that the total cross section $\sigma_T$ satisfies
\begin{equation}
c_1\mathrm{g}^{\rho+2}\leq \mathcal{F}\sigma_T(\mathrm{g}) \leq c_2\left( 1 + \mathrm{g}^{\rho+2} \right).
\label{Eq:GradCutOff3}
\end{equation}
Well-known models satisfying these bounds  include Israel particles and hard spheres in three dimensions (see Refs.~\cite{wI63} and \cite{CercignaniKremer-Book}) and hard disks in two dimensions (see Ref.~\cite{aGaM2019}). In these cases, the transport coefficients can be computed analytically.

Having discussed the assumptions on the cross section, one proceeds as follows. First, the linearized collision operator $\mathcal{L}[\phi]$ is written as the sum of three terms (known as the Hilbert decomposition):
\begin{equation}
\mathcal{L}[\phi] = \mathcal{M}_{\nu}[\phi] + \mathcal{K}_1[\phi] - \mathcal{K}_2[\phi].
\label{Eq:HilbertExpansionA}
\end{equation}
The first operator on the right-hand side is the multiplication operator $\mathcal{M}_\nu[\phi] := \nu_x\phi$ involving the function\footnote{Note that our function $\nu_x$ corresponds to $p^0$ times the collisional frequency $\nu(p)$ introduced in~\cite{mDmE88}.}
\begin{equation}
\nu_x(p_1) := \int\limits_{P_x^+(m)} k_1(p_1,p_2) d\mu_x(p_2),\qquad
k_1(p_1,p_2) :=  \mathcal{F}\sigma_T(\mathrm{g}) = m^2\mathrm{g}\sqrt{1 + \frac{\mathrm{g}^2}{4}}\sigma_T(\mathrm{g}),
\label{Eq:Defnux}
\end{equation}
whereas the operators $\mathcal{K}_1$ and $\mathcal{K}_2$ on the right-hand side of Eq.~\ref{Eq:HilbertExpansionA}) are given by
\begin{equation}
\mathcal{K}_1[\phi](p_1) := \int\limits_{P_x^+(m)}
 k_1(p_1,p_2)\phi(p_2) d\mu_x(p_2),
\label{Eq:K1Def}
\end{equation}
and
\begin{equation}
\mathcal{K}_2[\phi](p_1) := \int\limits_{P_x^+(m)}\int\limits_{S^{d-1}}
\mathcal{F}\frac{d\sigma}{d\Omega}(\mathrm{g},\Theta)
  \left[ \phi(p_1^*) + \phi(p_2^*) \right]
d\Omega(\hat{q}^*)d\mu_x(p_2),
\label{Eq:K2Def}
\end{equation}
respectively. Since the reflection $\hat{q}^*\mapsto -\hat{q}^*$ interchanges $p_1^*$ with $p_2^*$ but keeps $\mathcal{F}\frac{d\sigma}{d\Omega}$ invariant, the terms inside the square parenthesis on the right-hand side of Eq.~(\ref{Eq:K2Def}) yield equal contributions to the integral. Likewise, recalling the definition of the relative speed $\mathrm{g} := |p_2-p_1|/m$, it follows immediately that $k_1(p_1,p_2)$ is symmetric in $p_1$ and $p_2$.

Next, one analyses the properties of the multiplication operator. To this purpose, one first proves:

\begin{lemma}
\label{Lem:nu}
The function $\nu_x: P_x^+(m)\to \Real$ defined by Eq.~(\ref{Eq:Defnux}) is continuous and satisfies the following bounds. There are positive constants $C_2 > C_1 > 0$ such that for all $p\in P_x^+(m)$,
\begin{equation}
C_1\gamma^{\frac{\rho}{2}+1} \leq \nu_x(p) \leq C_2 \gamma^{\frac{\rho}{2}+1},\qquad
\gamma := \frac{-u^\mu p_\mu}{m}.
\label{Eq:nuEstimate}
\end{equation}
Since $\gamma\geq 1$ it follows, in particular, that $\nu_x$ has a positive lower bound and that its image consists of the interval $[\nu_1,\infty)$, where
\begin{equation}
\nu_1 := \inf\limits_{p\in P_x^+(m)} \nu_x(p)\geq C_1 > 0.
\end{equation}
\end{lemma}

An immediate corollary of Lemma~\ref{Lem:nu} is:

\begin{lemma}
\label{Lem:Collision1}
Define
\begin{equation}
D(\mathcal{L}) := \left\{ \phi\in {\cal H}_x : \nu_x\phi \in {\cal H}_x \right\}.
\end{equation}
Then, $\mathcal{M}_\nu: D(\mathcal{L}): {\cal H}_x\to {\cal H}_x$ is a densely-defined self-adjoint operator on ${\cal H}_x$ with spectrum $[\nu_1,\infty)$.
\end{lemma}

A proof of Lemma~\ref{Lem:nu}, which holds for arbitrary dimensions $d\geq 2$, is included in Appendix~\ref{App:ProofLemnu}.

The next step consists in proving that $\mathcal{K}_1$ and $\mathcal{K}_2$ are self-adjoint compact operators in ${\cal H}_x$. For $\mathcal{K}_1$ the statement follows immediately\footnote{See for instance~\cite{ReedSimon80I}, chapter~VI.} from the fact that $k(p_1,p_2)$ is real-valued, symmetric in $p_1$ and $p_2$ and that is satisfies
\begin{equation}
\int\limits_{P_x^+(m)}\int\limits_{P_x^+(m)} |k_1(p_1,p_2)|^2 d\mu_x(p_1) d\mu_2(p_2) < \infty,
\label{Eq:HilbertSchmidt}
\end{equation}
which is a direct consequence of the upper bound in Eq.~(\ref{Eq:GradCutOff3}), see Appendix~\ref{App:ProofLemnu} for more details. However, the corresponding statement for the operator $\mathcal{K}_2$ is much more elaborated. The first problem is to rewrite it in the form of an integral operator, similar to $\mathcal{K}_1$, such that
\begin{equation}
\mathcal{K}_2[\phi](p_1) = \int\limits_{P_x^+(m)}  k_2(p_1,p_1^*)\phi(p_1^*) d\mu_x(p_1^*),
\end{equation}
with a suitable integral kernel $k_2(p_1,p_1^*)$. This requires introducing new coordinates on the collision manifold and will be presented elsewhere. The second problem is to prove that $\mathcal{K}_2$ is compact. As far as we are aware, so far this has only been established in the three-dimensional case in~\cite{mDmE88}, based on the representation for $k_2(p_1,p_2^*)$ given in chapter~IX of Ref.~\cite{Groot-Book}. We hope to address the general case $d\geq 2$ in future work. Summarizing, one has

\begin{lemma}
\label{Lem:Collision2}
The integral operators $\mathcal{K}_1$ (for arbitrary $d\geq 2$) and $\mathcal{K}_2$ (for $d=3$) are compact and self-adjoint on $\mathcal{H}_x$.
\end{lemma}

In a final step, one invokes Weyl's theorem (see for instance chapter XIII.4 in ~\cite{ReedSimon80IV}) and concludes that $\mathcal{L}: D(\mathcal{L}): {\cal H}_x\to {\cal H}_x$ is self-adjoint. The inequality~(\ref{Eq:LinCollNonnegative})  implies that $\mathcal{L}$ is nonnegative and that its kernel is given by Eq.~(\ref{Eq:LinCollKernel}). Furthermore, Weyl's theorem implies that $\sigma(\mathcal{L})$ consists of $[\nu_1,\infty)$ and a sequence of eigenvalues in the interval $[0,\nu_1]$ which can only accumulate at $\nu_1$. In particular, there exists a smallest positive element $\nu_0$ of $\sigma(\mathcal{L})$ and
\begin{equation}
\langle \phi, \mathcal{L}[\phi] \rangle_x \geq \nu_0 \langle \phi,\phi \rangle_x,
\qquad
\phi\in D(\mathcal{L})\cap (\ker\mathcal{L})^\perp.
\end{equation}
This concludes our discussion regarding the mathematical treatment of the linearized collision operator.

\section{The first-order theory resulting from the Chapman-Enskog projection method}
\label{Sec:ChapmanEnskog}

In this section we implement the Chapmant-Enskog projection method, as described in Section~\ref{SubSec:CE}, invoking the results presented in Section~\ref{Sec:LinearizedCollisionTerm}. Accordingly, the main goal of this section is the determination of the first-order correction term $h_0$, as defined by Eq.~(\ref{h0}), and the derivation of the resulting fluid theory. However, before undertaking such task, it is worthwhile to recall that in order to obtain Eq.~(\ref{h0}) a particular frame was chosen. That is, the compatibility conditions corresponding to the trace-fixed particle frame, given in Eqs.~(\ref{Eq:TFPF1}) and (\ref{Eq:TFPF2}) were imposed on the distribution function in order to specify the state variables $n$, $T$, and $u^\mu$ parametrizing $f^{(0)}_\varepsilon$. Some relevant implications of such choice together with its non-relativistic limit are briefly discussed in Section~\ref{TFPFdepict} before explicitly obtaining the constitutive equations in Section~\ref{h0andConst.Eq.} and their invariant representations in Section~\ref{Sec:Invariant}. From here on we shall drop the index $\varepsilon$ and write $f^{(0)}$ and $\mathcal{L}$ instead of $f^{(0)}_\varepsilon$ and $\mathcal{L}_\varepsilon$ to simplify the notation.

\subsection{The trace-fixed particle frame}\label{TFPFdepict}

As mentioned above, the compatibility conditions specify the fields $n(x)$, $T(x)$, and $u^\mu(x)$ parametrizing the local J\"uttner distribution function~(\ref{Eq:f0nTu}). In particular, for the trace-fixed particle frame one imposes the conditions given by Eqs.~(\ref{Eq:TFPF1}) and (\ref{Eq:TFPF2}) within the Chapman-Enskog expansion $f_\varepsilon = f^{(0)} + \varepsilon h_\varepsilon$, which require that the first $d+2$ moments of $f_\varepsilon$ and $f^{(0)}$ coincide with each other. Moreover, they imply that $h_\varepsilon/f^{(0)}$ is orthogonal to $\ker\mathcal{L}$, which will turn out to be very useful for what follows.

Notice that, using Eq.~(\ref{Eq:Generatingf0}), one can write Eqs.~(\ref{Eq:TFPF1}) and (\ref{Eq:TFPF2}) as
\begin{equation}
J^\mu = \int\limits_{P_x^+(m)} 
 f(x,p) p^\mu\dvol_x(p) =nu^\mu,\qquad 
T^\mu_\mu = -m^2\int\limits_{P_x^+(m)} 
 f(x,p) \dvol_x(p) = -ne+d\mathrm{p} = -m n G_{-1}, 
\label{compatibilityBIS}
\end{equation}
and we now verify that these relations allow one to uniquely determine the fields $n(x)$, $z(x) = m/(k_B T(x))$, and $u^\mu(x)$ from $f^{(0)}$. Indeed, it is clear that Eq.~(\ref{Eq:TFPF2}) determines $n(x)$ and $u^\mu(x)$. Moreover, the fact that this choice leads to the particle current being aligned with the hydrodynamic velocity implies a particle frame.\footnote{Notice that we are considering as particle frames all of those in which there is no off-equilibrium contribution to the particle flux. This constitutes a whole family of frames given the freedom of still choosing how to fix the temperature. The Eckart frame, which is usually considered as \emph{the} particle frame, is only one of these frames, corresponding to fixing the temperature through the internal energy and thus imposing the matching condition $u_\mu u_\mu T^{\mu\nu}=ne$.} The temperature $T(x)$ is determined by means of Eq.~(\ref{Eq:TFPF1}) which corresponds to fixing the trace of the energy-momentum-stress tensor. This is possible since, by noticing that in the comoving frame $-u^\mu p_\mu = \sqrt{m^2 + |\vec{p}\,|^2}\geq m$, one can see that 
\[
\int\limits_{P_x^+(m)} 
 f^{(0)}(x,p) \dvol_x(p) \leq \frac{n(x)}{m}.
\]
Therefore, it follows from the invertibility of the function $G_{-1}$ established in Lemma~\ref{Lem:G-1} that Eq.~(\ref{Eq:TFPF1}) uniquely determines $z(x)$ and thus $T(x)$.

Before proceeding, it is instructive to work out the non-relativistic limit of the compatibility conditions~(\ref{Eq:TFPF1}) and (\ref{Eq:TFPF2}). For this, we assume there exists a local inertial frame in which most of the gas particles have $|\vec{p}\,|\ll m$. In this frame, the mean-particle velocity has the form $(u^\mu) = (1-|\vec{v}\,|^2)^{-1/2}(1,\vec{v}\,)$ with $|\vec{v}\,|\ll 1$ and one can write
\begin{equation}
p^0 = \sqrt{m^2 + |\vec{p}\,|^2}
 = m + \frac{|\vec{p}\,|^2}{2m} + {\cal O}\left( \frac{|\vec{p}\,|^4}{m^3} \right),
\qquad
\dvol_x(p) = \frac{d^d p}{p^0}.
\end{equation}
Hence, up to and including quadratic terms in the velocities, Eq.~(\ref{Eq:TFPF2}) gives
\begin{eqnarray}
&& \int f_\varepsilon(x,p) d^d p = \int f^{(0)}(x,p) d^d p = n(x)\left[ 1 + \frac{|\vec{v}(x)|^2}{2} \right],
\label{Eq:CompatibilityNewton1}\\
&&
\int \frac{\vec{p}}{m} 
 f_\varepsilon(x,p) d^d p = \int \frac{\vec{p}}{m} f^{(0)}(x,p) d^d p = n(x)\vec{v}(x),
\label{Eq:CompatibilityNewton2}
\end{eqnarray}
whereas Eq.~(\ref{Eq:TFPF1}), together with $Z = n G_{-1}/m$ and Eq.~(\ref{Eq:GkNonrel}), yields
\begin{equation}
\int \left( 1 - \frac{|\vec{p}\,|^2}{2m^2} \right) f_\varepsilon(x,p) d^d p = \int \left( 1 - \frac{|\vec{p}\,|^2}{2m^2} \right) f^{(0)}(x,p) d^d p = n\left[ 1 - \frac{d}{2m} k_B T \right].
\end{equation}
Combining the last equation with Eqs.~(\ref{Eq:CompatibilityNewton1}) and (\ref{Eq:CompatibilityNewton2}) leads to
\begin{equation}
\int \frac{|\vec{p}- m\vec{v}\,|^2}{2m} f_\varepsilon(x,p) d^d p = \int \frac{|\vec{p}- m\vec{v}\,|^2}{2m} f^{(0)}(x,p) d^d p = \frac{d}{2} n k_B T.
\label{Eq:CompatibilityNewton3}
\end{equation}
Hence, in the nonrelativistic limit, the compatibility conditions~(\ref{Eq:TFPF1}) and (\ref{Eq:TFPF2}) reduce to the usual ones used in the Newtonian theory. Indeed, the leading order in Eq.~ (\ref{Eq:CompatibilityNewton1}) defines the particle number density, Eq.~(\ref{Eq:CompatibilityNewton2}) fixes the hydrodynamic velocity, and (\ref{Eq:CompatibilityNewton3}) gives the internal energy as the average of the molecules' kinetic energy in the comoving frame~\cite{ChapmanCowling}.
 
\subsection{The first-order correction term $h_{0}$ and the constitutive equations}
\label{h0andConst.Eq.}

In order to determine the first-order correction $h_0$ to the local equilibrium distribution function and the expressions for the dissipative fluxes arising from it, here we consider Eq.~(\ref{h0}) with $f^{(0)}$ given by Eq.~(\ref{Eq:f0nTu}). For the right-hand side of Eq.~(\ref{h0}) we define
\begin{equation}
\mathcal{P}_{\mu\nu} := \frac{1}{k_B T}\left[ \sigma_{\mu\nu}-u_{\mu}\left(a_{\nu}+\frac{D_{\nu}T}{T}\right)+u_{\mu}u_{\nu}\left(\frac{\theta}{d}+\frac{\dot{T}}{T}\right) \right],
\label{Eq:PDef}
\end{equation}
such that
\begin{equation}
L_{F}[\log(Af^{(0)})] =	p^{\mu}p^{\nu}\mathcal{P}_{\mu\nu} + \frac{q}{k_B T} u^{\mu}p^{\nu}F_{\mu\nu}
 - \frac{m^2\theta}{d k_B T} + p^{\mu}\frac{\nabla_{\mu}\alpha}{\alpha},
\label{Eq:Lf0Decomposed1}
\end{equation}
where we have used $p^\mu p^\nu\Delta_{\mu\nu} = -m^2 + (p^\mu u_\mu)^2$. The next step consists in taking the orthogonal projection of the expression given in Eq.~(\ref{Eq:Lf0Decomposed1}). Since the last three terms on the right-hand side of Eq.~(\ref{Eq:Lf0Decomposed1}) belong to $\ker{\mathcal{L}}$, they project to zero, and one obtains
\begin{equation}
h_{0} = -f^{(0)} \mathcal{P}_{\mu\nu}\mathcal{L}^{-1}[(p^{\mu}p^{\nu})^{\perp}],
\label{Eq:h0inv}
\end{equation}
where we recall that $(p^\mu p^\nu)^\perp$ denotes the orthogonal projection of $p^\mu p^\nu$ onto $(\ker\mathcal{L})^\perp$. Although it will not be required for what follows, for completeness we provide its explicit form in Appendix~\ref{App:Projection}. Equation~(\ref{Eq:h0inv}) provides the complete first-order solution to the Boltzmann equation in the TFP frame. Recalling the expression for $T^{\mu\nu}$ given in Eq.~(\ref{Eq:Energy-Momentum-StressTensor}), the first-order correction to the energy-momentum-stress tensor yields
\begin{equation}
T^{\alpha\beta}(h_{0})= \int\limits_{P_x^+(m)} h_{0} p^\alpha p^\beta \dvol_x(p) = -\mathcal{P}_{\mu\nu}\mathcal{S}^{\mu\nu\alpha\beta},
\label{T1}
\end{equation}
where we have defined $\mathcal{S}^{\mu\nu\alpha\beta} := \langle\left(p^{\mu}p^{\nu}\right)^{\perp},\mathcal{L}^{-1}[(p^{\alpha}p^{\beta})^{\perp}]\rangle$. Using the fact that $\mathcal{L}^{-1}$ is symmetric and that $1\in \ker\mathcal{L}$, one recognizes that $\mathcal{S}^{\mu\nu\alpha\beta}$ has the properties (i) $\mathcal{S}^{\mu\nu\alpha\beta}=\mathcal{S}^{\alpha\beta\mu\nu}$,
(ii) $\mathcal{S}^{\mu\nu\alpha\beta}=\mathcal{S}^{\nu\mu\alpha\beta}=\mathcal{S}^{\nu\mu\beta\alpha}$,
(iii) $g_{\alpha\beta}\mathcal{S}^{\alpha\beta\mu\nu}=g_{\mu\nu}\mathcal{S}^{\alpha\beta\mu\nu} = 0$. Thus, it follows that it can be written in general as
\begin{align}
\mathcal{S}^{\alpha\beta\mu\nu} & =\mathcal{S}_{1}u^{\alpha}u^{\beta}u^{\mu}u^{\nu}+\mathcal{S}_{2}\Delta^{\alpha\beta}\Delta^{\mu\nu}+\frac{1}{2}\mathcal{S}_{3}\left(\Delta^{\alpha\mu}\Delta^{\beta\nu}+\Delta^{\beta\mu}\Delta^{\alpha\nu}\right)+\frac{1}{2}\mathcal{S}_{4}\left(\Delta^{\alpha\beta}u^{\mu}u^{\nu}+\Delta^{\mu\nu}u^{\alpha}u^{\beta}\right)\nonumber\\  
 & +\frac{1}{4}\mathcal{S}_{5}\left(\Delta^{\alpha\mu}u^{\beta}u^{\nu}+\Delta^{\alpha\nu}u^{\beta}u^{\mu}+\Delta^{\beta\mu}u^{\alpha}u^{\nu}+\Delta^{\beta\nu}u^{\alpha}u^{\mu}\right)\label{Sdecomp}
\end{align}
with $\frac{1}{d}\mathcal{S}_{1}=\frac{1}{2}\mathcal{S}_{4}=d\mathcal{S}_{2}+\mathcal{S}_{3}$. By substituting Eq.~(\ref{Sdecomp}) into Eq.~(\ref{T1}) and using the definition~(\ref{Eq:PDef}) one can write the first-order correction of $T_{\mu\nu}$ in the form
\begin{equation}
T_{\mu\nu}(\varepsilon h_{0})=  \epsilon^{(1)}\left( u_\mu u_\nu +\frac{1}{d} \Delta_{\mu\nu}\right)+ 2u_{(\mu} \mathcal{Q}^{(1)}\hspace{0cm}_{\nu)} + \mathcal{T}^{(1)}_{\mu\nu}
 \end{equation}
in accordance with Eq.~(\ref{Eq:epsilonTmunu}). Furthermore, the constitutive equations for the non-equilibrium quantities are given by 
\begin{equation}
\epsilon^{(1)}=-\varpi\left(\frac{\dot{T}}{T}+\frac{\theta}{d}\right),\qquad  \mathcal{Q}_{\mu}^{(1)}=-\lambda\left(\frac{D_{\mu}T}{T}+a_{\mu}\right),\qquad\mathcal{T}_{\mu\nu}^{(1)}=-2\eta\sigma_{\mu\nu},\label{Eq:ConstitutiveRelations}
\end{equation}
with the following expressions for the transport coefficients
\begin{equation}
\varpi=\varepsilon\frac{\mathcal S_1}{k_{B}T},\qquad
\lambda=\varepsilon\frac{\mathcal S_5}{4k_{B}T},\qquad
\eta=\varepsilon\frac{\mathcal S_3}{2k_{B}T}.\label{TransportCoefficientsTFP}
\end{equation}
The positiveness of these coefficients can be verified using the fact that $\mathcal{L}^{-1}$ is positive definite, since Eqs.~(\ref{T1}) and (\ref{Sdecomp}) allow one to write $\mathcal S_1$, $\mathcal S_3$, and $\mathcal S_5$ in terms of inner products  as follows
\begin{align}
\mathcal{S}_{1}&=\left\langle u_{\alpha}u_{\beta}\left(
p^{\alpha}p^{\beta}\right)^{\perp},\mathcal{L}^{-1}\left[u_{\mu}u_{\nu}\left(
p^{\mu}p^{\nu}\right)^{\perp}\right]\right\rangle,
\\
\mathcal{S}_{3}&=\frac{2}{\left(d-1\right)\left(d+2\right)}\left\langle \Delta_{\alpha\beta}\,^{\sigma\rho}\left(
p^{\alpha}p^{\beta}\right)^{\perp},\mathcal{L}^{-1}\left[\Delta_{\sigma\rho\mu\nu}\left(
p^{\mu}p^{\nu}\right)^{\perp}\right]\right\rangle, 
\\
\mathcal{S}_{5}&=\frac{4}{d}\left\langle \Delta_{\alpha\sigma}u_{\beta}\left(
p^{\alpha}p^{\beta}\right)^{\perp},\mathcal{L}^{-1}\left[\Delta_{\mu}^{\sigma}u_{\nu}\left(
p^{\mu}p^{\nu}\right)^{\perp}\right]\right\rangle, 
\end{align}
where $\Delta_{\alpha\beta}\,^{\mu\nu} := \Delta_{\alpha}\,^{\mu}\Delta_{\beta}\,^{\nu}-\frac{1}{d}\Delta_{\alpha\beta}\Delta^{\mu\nu}$. Moreover, following the transformations described in Appendix~A of Ref.~\cite{aGjSoS2024b}, one can write the three invariant transport coefficients $\eta$, $\zeta$, and $\kappa$ in terms of these quantities. Notice that the shear viscosity $\eta$ is always a frame and representation invariant coefficient, whereas in the TFP frame here considered the thermal conductivity yields $\kappa=\lambda$. In order to relate the coefficient $\varpi$ with the bulk viscosity, we first recall its invariant definition. If the first-order non-equilibrium contributions to the quantities $\mathcal{E}$ and $\mathcal{P}$ (see Appendix~\ref{App:Macroscopic}) are written in a general fashion as
\begin{align}
\epsilon^{(1)} &=
\varepsilon_1\frac{\dot{n}}{n} 
 + \varepsilon_2\frac{\dot{T}}{T} + \varepsilon_3\theta 
 ,
\label{Eq:ConstitutiveE}\\
\pi^{(1)} &=  
 \pi_1\frac{\dot{n}}{n} + \pi_2\frac{\dot{T}}{T} + \pi_3\theta,
\label{Eq:ConstitutiveP}
\end{align} 
the bulk viscosity $\zeta$ is given by ~\cite{pK19,aGjSoS2024b}
\begin{equation}
\zeta = \left( \pi_1 - \frac{k_B}{c_v}\varepsilon_1 \right)
 + \frac{k_B}{c_v} \left( \pi_2 - \frac{k_B}{c_v}\varepsilon_2 \right)
 - \left( \pi_3 - \frac{k_B}{c_v}\varepsilon_3 \right).
\end{equation}
Since in the TFP frame one has $\pi_1=\varepsilon_1=0$ and $\pi_i=\varepsilon_i/d$ with $\varepsilon_2=d\varepsilon_3=-\varpi$ for $i=2,3$, one obtains
\begin{equation}
\zeta=\varpi\left(\frac{1}{d}-\frac{k_{B}}{c_{v}}\right)^{2},
\label{zetamu}
\end{equation}
which is positive as well.

\subsection{First-order changes of frame and representation}
\label{Sec:Invariant}

As discussed above, the coefficients  which relate dissipative quantities with the first-order gradients through Eq.~(\ref{Eq:ConstitutiveRelations}) can be written in terms of the invariant transport coefficients $\eta$, $\zeta$, and $\kappa$. Moreover, as pointed out in Section~\ref{SubSec:CEVariants}, these equations can also be written in other frames, corresponding to the freedom of choice in the equilibrium state variables. These variants can be obtained from the phenomenological point of view by performing first-order transformations as discussed in Ref.~\cite{pK19}. In the presence of a weak electromagnetic field, the corresponding transformations are given in Ref.~\cite{aGjSoS2024b} where the particular case of the transformation between Eckart's frame and the TFP one is explicitly addressed. 

In the context of kinetic theory, as described in Section~\ref{SubSec:CEVariants}, these changes correspond to the addition of a linear combination of the elements of $\ker\mathcal{L}$ to the right-hand side of Eq.~(\ref{Eq:h0inv}). In general, this step leads to a family of solutions and uniqueness is then achieved by fixing the coefficients involved through the enforcement of the compatibility conditions (see for example Refs.~\cite{ChapmanCowling,CercignaniKremer-Book}). As pointed out in Section~\ref{SubSec:CE}, the particular choice of the TFP frame has no such contribution. In order to explicitly prove this statement and further illustrate this point by obtaining the constitutive equations corresponding to Eckart's frame, we consider, for the first-order off equilibrium contribution to the distribution function
\begin{equation}
f^{(1)} = -f^{(0)}\left\{ \mathcal{P}_{\mu\nu}\mathcal{L}^{-1}[(p^{\mu}p^{\nu})^{\perp}] + m A + B_{\mu}p^{\mu}\right\},
\label{eq:h0inv}
\end{equation}
where $A$ and $B^{\mu}$ are arbitrary functions of the state variables, which are to be determined through the compatibility conditions. These conditions can be written in a general fashion as
\begin{align}
\int\limits_{P_{x}^{+}(m)}g_{1}(\gamma)f^{(1)}(x,p)\dvol_{x}(p) & =0,\label{eq:matchgeneral1}\\
\int\limits_{P_{x}^{+}(m)}g_{2}(\gamma)f^{(1)}(x,p)\dvol_{x}(p) & =0,\label{eq:matchgeneral2}\\
\int\limits_{P_{x}^{+}(m)}g_{3}(\gamma)\Delta_{\mu\nu}p^{\nu}f^{(1)}(x,p)\dvol_{x}(p) & =0.\label{eq:matchgeneral3}
\end{align}
where $g_{1}$, $g_{2}$, and $g_{3}$ are three arbitrary functions ($g_1$ and $g_2$ being independent from each other)  and we have introduced the variable $\gamma:=-u_{\mu}p^{\mu}/m$ as in Eq.~(\ref{Eq:nuEstimate}) in order to simplify the notation. Notice that for a particle frame one considers $g_{2}(\gamma)=\gamma$ and $g_{3}(\gamma)=1$, leading to Eq.~(\ref{Eq:TFPF2}). Within this choice, the trace-fixed condition in Eq.~(\ref{Eq:TFPF1}) corresponds to $g_{1}(\gamma)=1$, while Eckart's frame is recovered for $g_{1}(\gamma)=\gamma^{2}$. Moreover, it is easy to check that the general matching conditions in terms of the three indices $(q,s,z)$ considered in Eqs.~(17) and (18) of Ref.~\cite{Rocha2022} correspond to $g_{1}(\gamma)=\gamma^{s}$, $g_{2}(\gamma)=\gamma^{q}$, and $g_{3}(\gamma)=\gamma^{z}$.

Considering Eq.~(\ref{eq:h0inv}) and decomposing $B_\mu = b_\mu - b u_\mu$ with $b_\mu:=\Delta_\mu{}^\nu B_\nu$, one can derive constitutive equations for the dissipative components of $J^{\alpha}$ and $T^{\alpha\beta}$ in an arbitrary frame. A closer study of this point is beyond the scope of this work and will be published elsewhere~\cite{aGaMoS26}. Here, we limit ourselves to an illustrative example. For a similar reasoning the reader is referred to Ref.~\cite{rHpK22} within Hilbert's method and Refs.~\cite{Rocha2022,Rocha_2023} within a modified Chapman-Enskog approach.

For the particular case of a particle frame, one obtains $b_{\mu}=0$ and $b=-\frac{m}{e}A$ with
\begin{equation}
A = -\frac{1}{m}\frac{\left\langle g_{1}(\gamma),\mathcal{L}^{-1}[(p^{\mu}p^{\nu})^{\perp}]\right\rangle \mathcal{P}_{\mu\nu}}{\mathcal{W}_{0,0}[g_{1}(\gamma)f^{(0)}] - \frac{m}{e}\mathcal{W}_{1,0}[g_{1}(\gamma)f^{(0)}]},
\label{Aparticle}
\end{equation}
where the functionals $\mathcal{W}_{n,k}[\phi]$ are defined in Appendix \ref{App:GeneralMomentIntegrals}. Notice that these results consistently lead to $\mathcal{J}^\mu=0$ and $\mathcal{N}=n$ respectively (see Eq.~(\ref{Eq:generalJ})). In the particular case of the TFP frame, the values $A=b=0$ are verified by the previous expressions (since $g_{1}\left(\gamma\right)=1$ which lies in $\ker\mathcal{L}$) and thus the homogeneous solution does not contribute to the dissipative fluxes, as expected. However, if one considers Eckart's frame, for which $g_{1}\left(\gamma\right)=\gamma^{2}$,
the non-vanishing contribution arising from the second and third terms on the right-hand side of Eq.~(\ref{eq:h0inv}) leads to constitutive relations different from the ones in Eq.~(\ref{Eq:ConstitutiveRelations}). With $g_1(\gamma)=\gamma^2$ one obtains the following result for $A$ and $b$: 
\begin{equation}
A=-\frac{e}{m}b=\frac{k_{B}}{c_{v}}\frac{ez^{2}}{nm^{3}}\frac{\mathcal{S}_{1}}{k_{B}T}\left(\frac{\dot{T}}{T}+\frac{\theta}{d}\right),
\end{equation}
which leads to
\begin{align}
\epsilon^{(1)}_E=u_{\alpha}u_{\beta}T^{\alpha\beta}(\varepsilon h_0)&=0,
\\
\pi_E^{(1)}=\frac{1}{d}\Delta_{\alpha\beta}T^{\alpha\beta}(\varepsilon h_0)&=-\varpi\left(\frac{1}{d}-\frac{k_{B}}{c_{v}}\right)\left(\frac{\dot{T}}{T}+\frac{\theta}{d}\right),
\label{Eq:piE}
\end{align}
where $\varpi$ is the same transport coefficient as the one obtained in the TFP frame (see Eq. (\ref{TransportCoefficientsTFP})). 

Notice that the constitutive relation for the pressure deviator $\pi_E^{(1)}$ given in Eq.~(\ref{Eq:piE}) does not match the phenomenological result that one obtains by enforcing the second law of thermodynamics in Eckart's frame, namely \cite{cE1940},
\begin{align}
\qquad\mathcal{Q}_{\mu}^{(1)}=-\kappa\left(\frac{D_{\mu}T}{T}+a_{\mu}\right),\qquad\pi_E^{(1)}=-\zeta\theta,\qquad\mathcal{T}_{\mu\nu}^{(1)}=-2\eta\sigma_{\mu\nu}.
\label{Eq:EckartConstitutiveRelations}
\end{align}
However, one can still exploit the additional freedom described in Section~\ref{SubSec:CEVariants} in order to change the representation. Indeed, by adding the expression in Eq.~(\ref{Eq:repchange}) to the right-hand side of Eq.~(\ref{Eq:CEImplicitProjected}) one is led to $h_{0} = - f^{(0)}\widehat{\mathcal{P}}_{\mu\nu}\mathcal{L}^{-1}[(p^{\mu}p^{\nu})^{\perp}]$ with
\begin{align}
\widehat{\mathcal{P}}_{\mu\nu} & := \frac{1}{k_{B}T}\left\{ \sigma_{\mu\nu} - u_{\mu}\left[a_{\nu}+\frac{D_{\nu}T}{T}-\frac{\hat{\Gamma}_{2}}{k_{B}T}\left(h a_{\nu}+\frac{D_{\nu}p}{n} - qF_{\nu\alpha} u^{\alpha} \right)\right]\right.
\nonumber \\
 & \left.+\left[\frac{\dot{T}}{T}+\frac{\theta}{d} + \hat{\Gamma}_{0}\left(\frac{\dot{n}}{n}+\theta\right) - \hat{\Gamma}_{1}\left(\frac{c_{v}}{k_{B}}\frac{\dot{T}}{T}+\theta\right) 
 \right]u_{\mu}u_{\nu} \right\},
\label{Eq:Phatmunu}
\end{align}
where the zeroth-order balance equations have been introduced. By choosing $\hat{\Gamma}_1 = k_B/c_v$ and $\hat{\Gamma}_2 = 0$ one recovers the constitutive relations~(\ref{Eq:EckartConstitutiveRelations}) in Eckart's frame.

In the TFP frame, the modification~(\ref{Eq:Phatmunu}) leads to
\begin{align}
\epsilon^{(1)}&=-\varpi\left[\frac{\dot{T}}{T}+\frac{\theta}{d}-\frac{c_{v}}{k_{B}}\hat{\Gamma}_{1}\left(\frac{\dot{T}}{T}+\frac{k_{B}}{c_{v}}\theta\right)\right],
\label{Eq:ConstBuenas1}\\
\mathcal{Q}_{\mu}^{(1)}&=-\kappa\left[\frac{D_{\mu}T}{T} + a_{\mu} - \frac{h}{k_{B}T}\hat{\Gamma}_{2}\left(a_{\mu} + \frac{D_{\mu}p}{nh} - \frac{q}{h}F_{\mu\nu}u^{\nu}\right)\right],
\label{Eq:ConstBuenas2}\\
\mathcal{T}_{\mu\nu}^{(1)} &= -2\eta\sigma_{\mu\nu},
\label{Eq:ConstBuenas3}
\end{align}
where  $\hat{\Gamma}_{1,2}$ are arbitrary functions of $n$ and $T$. The freedom of choosing the values of these parameters within an already fixed frame allows for a change in the force-flux relations which only alters second-order terms but has a substantial impact on the properties of the system of transport equations. In fact, as follows from Eqs.~(\ref{Eq:HeatDiss}) and (\ref{Eq:EulerDiss}), the terms multiplying $\hat{\Gamma}_{1}$ and $\hat{\Gamma}_{2}$ are of order $\varepsilon k^2$ (whereas the remaining ones are of order $k$), such that they do not affect the first-order content of the theory.

It is worthwhile to comment at this point that, for any choice of $g_1$, $g_2$, and $g_3$, the addition of the homogeneous solution cannot change the structure of the force-flux relations. As mentioned above, this is thoroughly discussed in Ref.~\cite{aGaMoS26}; however it can be foreseen by inspecting Eqs.~(\ref{eq:h0inv}--\ref{eq:matchgeneral3}). Indeed, the general conditions in Eqs.~(\ref{eq:matchgeneral1})-(\ref{eq:matchgeneral3}) lead to values for $A$, $b$, and $b_\mu$ containing linear combinations of $\mathcal {P}_{\mu\nu}$ in such a way that combinations of time or space derivatives of state variables others than the ones appearing in $\mathcal{P}_{\mu\nu}$ cannot be introduced with this freedom. Therefore, a change of frame can only lead to different coefficients in the constitutive relations but the force-flux structure is retained. However, notice that it is possible that in some frames, one or several coefficients vanish as is the case in the example provided above.

The system of partial differential equations resulting from introducing the constitutive relations (\ref{Eq:ConstBuenas1}--\ref{Eq:ConstBuenas3}) in Eqs.~(\ref{Eq:ContinuityDiss}--\ref{Eq:EulerDiss}) leads to the system proposed in Refs.~\cite{aGjSoS2024a} and \cite{aGjSoS2024b} where the parameters $\Gamma_{1,2}$ in these references are related to $\hat{\Gamma}_{1,2}$ according to $\hat{\Gamma}_1 = \frac{k_B}{c_v}\Gamma_1$ and $\hat{\Gamma}_2 = \frac{k_B T}{h}\Gamma_2$. As proven in Refs.~\cite{aGjSoS2024a,aGjSoS2024b}, the resulting system is hyperbolic, causal, and stable at global equilibrium states, provided that $\Gamma_1$ is chosen large enough, $\Gamma_2 = h/(k_B T)$ (which corresponds to $\hat{\Gamma}_2 = 1$), and the transport coefficients satisfy the inequality $1 + 2\eta/\kappa \le e/(k_B T)$.  Moreover, the resulting nonlinear system of equations, together with suitable initial data, possesses a (local in time) well-posed Cauchy problem, see Theorem~1 in~\cite{aGjSoS2024b}.

\section{Entropy current density and second law of thermodynamics}
\label{Sec:entropy}

This section is dedicated to a brief discussion regarding the form of the entropy current in the Chapman-Enskog approach and on the validity of the second law of thermodynamics. Of course, Boltzmann's H-theorem guarantees a nonnegative entropy production, as discussed in Section~\ref{Sec:Preliminaries}. However, when truncating the Chapman-Enskog series to first order in the Knudsen parameter $\varepsilon$, as we do in this article, the corresponding entropy current density is not guaranteed to have strict nonnegative divergence. Nevertheless, as we show in this section, the leading-order terms in the entropy production, which are quadratic in $\varepsilon$, are positive-definite. In this sense, the second law of thermodynamics holds within the range of validity of the first-order theory.

In order to compute the entropy current density defined in Eq.~(\ref{Eq:EntropyCurrentDensity}) we first recast the Chapman-Enskog ansatz~(\ref{Eq:CEAnsatz}) into the form
\begin{equation}
f_\varepsilon = f_\varepsilon^{(0)}\left( 1 + \varepsilon\frac{h_\varepsilon}{f_\varepsilon^{(0)}} \right),
\end{equation}
which yields
\begin{equation}
f_\varepsilon\log(A f_\varepsilon) = f_\varepsilon^{(0)}\log(A f_\varepsilon^{(0)}) + \varepsilon\left[ \log(A f_\varepsilon^{(0)}) + 1 \right] h_\varepsilon
 + \varepsilon^2\sum\limits_{j=0}^\infty\frac{(-\varepsilon)^j}{(j+1)(j+2)}\left( \frac{h_\varepsilon}{f_\varepsilon^{(0)}} \right)^{j+2} f_\varepsilon^{(0)},
\end{equation}
with the series converging absolutely if $|h_\varepsilon|\leq f_\varepsilon^{(0)}$. From Eq.~(\ref{Eq:f0Bis}) we have
\begin{equation}
\log(A f_\varepsilon^{(0)}) + 1 = \frac{\mu + u^\alpha p_\alpha}{k_B T},
\end{equation}
where for notational simplicity we have omitted the subscripts $\varepsilon$ in $\mu$, $T$, and $u^\alpha$. From this, we obtain the following expansion of the entropy current density:
\begin{equation}
S^\alpha(f_\varepsilon) = S^\alpha(f_\varepsilon^{(0)}) - \varepsilon \frac{\mu J^\alpha(h_\varepsilon) + T^{\alpha\beta}(h_\varepsilon) u_\beta}{T} - k_B\varepsilon^2\sum\limits_{j=0}^\infty\frac{(-\varepsilon)^j}{(j+1)(j+2)} J^\alpha\left( \frac{h_\varepsilon^{j+2}}{(f_\varepsilon^{(0)})^{j+1}} \right),
\end{equation}
where the zeroth-order contribution is
\begin{equation}
S^\alpha(f_\varepsilon^{(0)}) 
 = -\frac{(\mu-k_B T) J^\alpha(f_\varepsilon^{(0)}) + T^{\alpha\beta}(f_\varepsilon^{(0)}) u_\beta}{T}
 = n s u^\alpha,\qquad s = \frac{h - \mu}{T}.
\end{equation}
Therefore, including first-order corrections, we obtain
\begin{equation}
S^\alpha(f_\varepsilon) = S_{IS}^\alpha(f_\varepsilon) 
 + \mathcal{O}(\varepsilon^2),
\label{Eq:ISEntropy}
\end{equation}
with $S_{IS}^\alpha$ being the Israel-Stewart entropy current density  defined as~\cite{wIjS79a,Israel1987} 
\begin{equation}
S_{IS}^\alpha(f_\varepsilon) :=
\frac{1}{T}\left[ \mathrm{p} u^\alpha - \mu J^\alpha(f_\varepsilon) - T^{\alpha\beta}(f_\varepsilon) u_\beta \right]. 
\end{equation}
Of course, one can replace $f_\varepsilon$ with $f_\varepsilon^{(0)} + \varepsilon h_0$ in $S_{IS}^\alpha$ and Eq.~(\ref{Eq:ISEntropy}) is still correct. From now on, we assume that we evaluate $S_{IS}^\alpha$ using the latter choice and omit the argument for notational simplicity.

As reviewed in Appendix~B of Ref.~\cite{aGjSoS2024b}, the Israel-Stewart entropy current satisfies the covariant Gibbs relation
\begin{equation}
\delta S_{IS}^\alpha = -\frac{u_\beta}{T}\delta T^{\alpha\beta} - \frac{\mu}{T}\delta J^\alpha,
\label{Eq:CovariantGibbsRel}
\end{equation}
for arbitrary perturbations around an equilibrium configuration. Consequently, $S_{IS}^\alpha$ is first-order frame invariant. Furthermore, using the balance laws and the notation in Appendix~\ref{App:Macroscopic} one obtains
\begin{align}
T\nabla_\alpha S_{IS}^\alpha &= -(\mathcal{E} - n e)\frac{\dot{T}}{T} 
- (\mathcal{P} - \mathrm{p})\theta - (\mathcal{N} - n)T\overset{\cdot}{\left( \frac{\mu}{T} \right)}
\nonumber\\
 &- \left( \mathcal{Q}^\alpha - h\mathcal{J}^\alpha \right)\left( \frac{D_\alpha T}{T} + a_\alpha \right) - h\mathcal{J}^\alpha\left( a_\alpha + \frac{D_\alpha \mathrm{p}}{nh} - \frac{q}{h} E_\alpha \right) - \mathcal{T}^{\alpha\beta}\sigma_{\alpha\beta},
\label{Eq:EntropyProd1}
\end{align}
which reduces to Eq.~(B5) in~\cite{aGjSoS2024b} in a particle frame, for which $\mathcal{N} = n$ and $\mathcal{J}^\alpha=0$. In particular, for the constitutive relations~(\ref{Eq:ConstitutiveRelations}) in the TFP frame derived in the previous section, one finds
\begin{equation}
T\nabla_\alpha S_{IS}^\alpha = \varpi\left( \frac{\dot{T}}{T} + \frac{\theta}{d} \right)^2 + \lambda\left( \frac{D^\alpha T}{T} + a^\alpha \right)\left( \frac{D_\alpha T}{T} + a_\alpha \right) + 2\eta\sigma^{\alpha\beta}\sigma_{\alpha\beta},
\label{Eq:EntropyProd2}
\end{equation}
which is positive-definite, given that the transport coefficients $\varpi$, $\lambda$, and $\eta$ are positive. Therefore, the theory associated with the constitutive equations~(\ref{Eq:ConstitutiveRelations}) has exact nonnegative entropy production! However, as discussed towards the end of the last section, in order to achieve hyperbolic evolution equations in the resulting fluid theory, one needs to consider the modified constitutive relations~(\ref{Eq:ConstBuenas1}--\ref{Eq:ConstBuenas3}). This leads to additional terms on the right-hand side of Eq.~(\ref{Eq:EntropyProd2}) and the resulting entropy production is not manifestly nonnegative anymore. However, these additional terms are third-order in the gradients, and hence they constitute higher-order terms. In this sense, the second law of thermodynamics still holds within the limit of validity of our first-order theory.

We end this section with two comments. The first one refers to the decomposition $S_{IS}^\alpha = \mathcal{S} u^\alpha + \phi^\alpha$ of the Israel-Stewart entropy current for which (in the notation of Appendix~\ref{App:Macroscopic}),
\begin{align}
\mathcal{S} &= n s + \frac{\mathcal{E} - n e}{T} - (\mathcal{N}-n)\frac{\mu}{T},
\label{Eq:SIS}\\
\phi^\alpha &= \frac{\mathcal{Q}^\alpha - \mu\mathcal{J}^\alpha}{T},
\label{Eq:phiIS}
\end{align}
such that one can rewrite Eq.~(\ref{Eq:EntropyProd1}) in the form
\begin{align}
T\nabla_\alpha S_{IS}^\alpha &= -(\mathcal{S} - n s)\dot{T} - (\mathcal{N} - n)\dot{\mu} - (\mathcal{P} - \mathrm{p})\theta 
\nonumber\\
 &- T\phi^\alpha\left( \frac{D_\alpha T}{T} + a_\alpha \right) - h\mathcal{J}^\alpha\left( a_\alpha + \frac{D_\alpha \mathrm{p}}{nh} - \frac{q}{h} E_\alpha \right) - \mathcal{T}^{\alpha\beta}\sigma_{\alpha\beta}.
\label{Eq:EntropyProd3}
\end{align}

The second comment refers to the comparison of our results with the Eckart frame, for which $\mathcal{E} = ne$ in addition to $\mathcal{N}=n$ and $\mathcal{J}^\alpha = 0$. In this case, Eq.~(\ref{Eq:SIS}) yields $\mathcal{S} = ns$, and thus it follows either from Eq.~(\ref{Eq:EntropyProd1}) or Eq.~(\ref{Eq:EntropyProd3}) and the constitutive relations $\mathcal{P} = \mathrm{p} - \zeta\theta$, $\mathcal{Q}^\alpha = -\kappa(D^\alpha T/T + a^\alpha)$, and $\mathcal{T}^{\alpha\beta} = -2\eta\sigma^{\alpha\beta}$ that
\begin{equation}
T\nabla_\alpha S_{IS}^\alpha = \zeta\theta^2 + \kappa\left( \frac{D^\alpha T}{T} + a^\alpha \right)\left( \frac{D_\alpha T}{T} + a_\alpha \right) + 2\eta\sigma^{\alpha\beta}\sigma_{\alpha\beta},
\label{Eq:EntropyProd4}
\end{equation}
which is also positive-definite. The resulting theory is not hyperbolic; however it can be made hyperbolic by changing the frame and representation (again at the cost of introducing third-order terms in the entropy production), see~\cite{Bemfica2022}. Note that the fact that $\mathcal{E} = n e$ and $\mathcal{S} = ns$ imply that in the Eckart frame the definition of the inverse temperature by means of the statistical formula
\begin{equation}
\frac{1}{T} = \left. \frac{\partial s}{\partial e} \right|_n,
\end{equation}
is valid beyond equilibrium including first-order corrections. In this sense, $T$ deserves the name ``temperature'' also for off-equilibrium configurations in the Eckart theory. In contrast, in the TFP frame, neither $\mathcal{E} - n e$ nor $\mathcal{S} - ns$ are zero off equilibrium and thus $T$ should be regarded as a ``temperature parameter" in this case.

\section{Conclusions}
\label{Sec:Conclusions}

In this article we provided a detailed microscopic derivation of the first-order theory for relativistic dissipative fluids put forward and shown to be hyperbolic, causal, and stable in Refs.~\cite{aGjSoS2024a,aGjSoS2024b}. Our derivation is based on a Chapman-Enskog approximation for the near-equilibrium solutions of the general relativistic Boltzmann equation. In contrast to previous works (see for example~\cite{wI63,Groot-Book,CercignaniKremer-Book,Rocha2022}), we have developed a new scheme within this approximation which follows the lines of~\cite{LaurePaper,LaureLectureNotes,JNET24} and is based on inverting the linearized collision operator on the orthogonal complement of its kernel in a suitable Hilbert space. The selection of this space guarantees the existence and uniqueness of the solution  at each order. Furthermore, as we have discussed, this space leads naturally to the trace-fixed particle frame adopted in~\cite{aGjSoS2024a,aGjSoS2024b}. However, as explained in this work, the resulting theory can be translated to any hydrodynamic frame.

Moreover, we have demonstrated that the force-flux representation-freedom (i.e. the freedom to add combinations of the Euler equations to the constitutive relations) can be incorporated in our Chapman-Enskog scheme by adding a suitable term to the Boltzmann equation which is zero on-shell. As explained in~\cite{aGjSoS2024a,aGjSoS2024b} exploiting this representation-freedom is fundamental to achieve hyperbolicity and stability of global equilibria, and it also constitutes a key ingredient in BDNK theories~\cite{Bemfica2022,pKovtun19} which are based on more general frames.

Although only applied here to derive the first-order theory, our scheme can in principle also be used to derive a fluid theory to any desired order of the Knudsen parameter. It should be interesting to investigate whether the second- or higher-order approximations obtained in this way, together with the representation-freedom, still lead to physically sound fluid theories.

We hope that this work provides the foundations for a new understanding of the Chapman-Enskog approximation in the relativistic regime and its applications to derive physically sound equations of relativistic dissipative fluids. It should be interesting to generalize it to  consider more general situations including mixtures of gases, an investigation of the strong field hydrodynamic regime, or the extension to spinning particles (see for example Refs.~\cite{Weickgenannt:2021cuo,Das:2022azr}).

\acknowledgments

We thank Guillermo Chac\'on, Alma R. M\'endez, Oscar Reula, J. F\'elix Salazar, and Thomas Zannias for fruitful comments and discussions throughout this work. We acknowledge support from SECIHTI through grant No. CBF-2025-G-1626. O.S. was partially supported by CIC grant No.~18315 to Universidad Michoacana de San Nicolás de Hidalgo. O.S. also thanks Universidad Aut\'onoma Metropolitana-Cuajimalpa, where this work was initiated, for hospitality.

\appendix

\section{Balance laws for the macroscopic quantities}
\label{App:Macroscopic}

In this appendix we provide a brief summary of the balance laws for the macroscopic quantities on a fixed background spacetime $(M,g)$ and electromagnetic field $F$. These laws consist of the following relations involving the particle current density vector field $J^\mu$, the energy-momentum-stress tensor $T^{\mu\nu}$, and the entropy current density vector field $S^\mu$,
\begin{equation}
\nabla_\mu J^\mu = 0,\qquad
\nabla_\mu T^{\mu\nu} + q J_\mu F^{\mu\nu} = 0,\qquad
\nabla_\mu S^\mu\geq 0.
\label{Eq:DivergenceLaws}
\end{equation}
Given an arbitrary future-directed timelike vector field $u^\mu$, normalized such that $u^\mu u_\mu = -1$, and the corresponding projection $\Delta^\mu{}_\nu = \delta^\mu{}_\nu + u^\mu u_\nu$ one may decompose these fields according to
\begin{eqnarray}
J^\mu &=& \mathcal{N} u^\mu + \mathcal{J}^\mu,\label{Eq:generalJ}\\
T^{\mu\nu} &=& \mathcal{E} u^\mu u^\nu + 2u^{(\mu}\mathcal{Q}^{\nu)} 
 + \mathcal{P}\Delta^{\mu\nu} + \mathcal{T}^{\mu\nu},\\
S^\mu &=& \mathcal{S} u^\mu + \phi^\mu,
\\
F^{\mu\nu} &=& 2 u^{[\mu} E^{\nu]} + \mathcal{B}^{\mu\nu}.\label{Eq:generalJT}
\end{eqnarray}
Here, $\mathcal{J}^\mu$, $\mathcal{Q}^\mu$, $\phi^\mu$, $E^\mu$, $\mathcal{T}^{\mu\nu}$, and $\mathcal{B}^{\mu\nu}$ are orthogonal to $u^\mu$ with $\mathcal{T}^{\mu\nu}$ being symmetric and trace-less and $\mathcal{B}^{\mu\nu}$ being antisymmetric. The square brackets denote the antisymmetric part of a tensor field. With respect to an observer that is comoving with the velocity field $u^\mu$, $\mathcal{N}$, $\mathcal{E}$, and $\mathcal{S}$ are the particle, energy, and entropy densities and $\mathcal{P}$ is the total pressure. Further, $\mathcal{J}^\mu$, $\mathcal{Q}^\mu$, and $\phi^\mu$ are the particle drift, heat flow, and entropy flow and $\mathcal{T}^{\mu\nu}$ denotes the trace-free part of the viscous tensor. Finally, $E^\mu = F^{\mu\nu} u_\nu$ is the electric vector field and $\mathcal{B}^{\mu\nu}$ the magnetic part.\footnote{In $d=3$ spatial dimensions $\mathcal{B}_{\mu\nu} = \varepsilon_{\mu\nu\alpha\beta} u^\alpha B^\beta$ is dual to the magnetic (pseudo-) vector field $B^\mu$.}

In terms of these quantities, the balance laws~(\ref{Eq:DivergenceLaws}) yield
\begin{eqnarray}
&& \dot{\mathcal{N}} + \theta\mathcal{N} + (D_\mu + a_\mu)\mathcal{J}^\mu = 0,
\label{Eq:FluidContinuityEq}\\
&& \dot{\mathcal{E}} + \theta(\mathcal{E} + \mathcal{P}) + (D_\mu + 2a_\mu)\mathcal{Q}^\mu + \sigma_{\mu\nu}\mathcal{T}^{\mu\nu} - q\mathcal{J}_\mu E^\mu = 0,
\label{Eq:FluidEnergyEq}\\
&& (\mathcal{E} + \mathcal{P}) a^\nu + D^\nu\mathcal{P}  + \dot{\mathcal{Q}}^\nu + \frac{d+1}{d}\theta\mathcal{Q}^\nu + (\sigma^{\mu\nu} + \omega^{\mu\nu})\mathcal{Q}_\mu + (D_\mu + a_\mu)\mathcal{T}^{\mu\nu} - q\left( \mathcal{N} E^\nu - \mathcal{J}_\mu\mathcal{B}^{\mu\nu} \right) = 0,\qquad
\label{Eq:FluidEulerEq}\\
&& \dot{\mathcal{S}} + \theta\mathcal{S} + (D_\mu + a_\mu)\phi^\mu \geq 0,
\label{Eq:FluidEntropyLaw}
\end{eqnarray}
where $a^\mu := u^\alpha\nabla_\alpha u^\mu$, $\theta := \nabla_\mu u^\mu$, $\sigma_{\mu\nu} := \left( \Delta_{(\mu}{}^\alpha\Delta_{\nu)}{}^\beta - d^{-1}\Delta_{\mu\nu}\Delta^{\alpha\beta} \right) \nabla_\alpha u_\beta$, and $\omega_{\mu\nu} := \Delta_{[\mu}{}^\alpha\Delta_{\nu]}{}^\beta\nabla_\alpha u_\beta$ refer to the acceleration, expansion, shear, and vorticity, respectively, associated with $u^\mu$. Further, the dot and the operator $D_\alpha$ refer to the application of the operators $u^\alpha\nabla_\alpha$ and $\Delta_{\alpha}{}^\beta\nabla_\beta$ and the projection of all free indices with $\Delta$. For example, $\dot{\mathcal{Q}}^\mu = \Delta^\mu{}_\nu u^\alpha\nabla_\alpha\mathcal{Q}^\nu$ and $D_\mu\mathcal{Q}^\mu = \Delta_{\mu\nu}\nabla^\mu\mathcal{Q}^\nu$, see Appendix~C in Ref.~\cite{aGjSoS2024b} for more details.

Local equilibrium configurations are characterized by the conditions
\begin{equation}
\mathcal{J}^\mu = 0,\quad
\mathcal{Q}^\mu = 0,\quad \mathcal{T}^{\mu\nu} = 0,\quad \phi^\mu = 0,
\end{equation}
such that $J^\mu$, the timelike eigenvector of $T^\mu{}_\nu$, and $S^\mu$ all point in the same direction. In this case, we write
\begin{equation}
\mathcal{N} = n,\quad
\mathcal{E} = n e,\quad
\mathcal{P} = \mathrm{p},\quad
\mathcal{S} = n s,
\end{equation}
with $n$, $e$, $\mathrm{p}$, and $s$ denoting the particle density, internal energy per particle, hydrostatic pressure, and entropy per particle, respectively. In this case, Eqs.~(\ref{Eq:FluidContinuityEq}--\ref{Eq:FluidEntropyLaw}) reduce to
\begin{eqnarray}
&& \dot{n} + \theta n = 0,
\label{Eq:FluidContinuityEqui}\\
&& \dot{e} + \theta\frac{\mathrm{p}}{n} = 0,
\label{Eq:FluidEnergyEqui}\\
&& h a^\nu + \frac{1}{n} D^\nu \mathrm{p} = q E^\nu,\qquad
h = e + \frac{\mathrm{p}}{n},
\label{Eq:FluidEulerEqui}\\
&& \dot{s}\geq 0.
\end{eqnarray}

\section{Recursion relations and other properties of the functions $G_k$}
\label{App:PropertiesGFunction}

In this appendix, we discuss the most important properties of the functions $G_k$ which are used throughout this work. Recall their  definition from Eq.~(\ref{Eq:Gk}),
\begin{equation}
G_k(z) := \frac{\textbf{K}_{\frac{d+1}{2} + k}(z)}{\textbf{K}_{\frac{d+1}{2}}(z)},\qquad
k = -1,0,1,2,\ldots,
\end{equation}
where $\textbf{K}_\nu$ refers to the modified Bessel functions of the second kind (see~\cite{DLMF-Book,DLMF} for definitions and properties). The identities
\begin{eqnarray}
    \label{Eq:Identity01}
    z^\nu \frac{d}{dz}\left[ z^{-\nu} \textbf{K}_\nu(z) \right] &=& -\textbf{K}_{\nu+1}(z), \\
    \label{Eq:Identity02}
    \textbf{K}_{\nu+1}(z) &=& \textbf{K}_{\nu-1}(z) + \frac{2\nu}{z}\textbf{K}_{\nu}(z),
\end{eqnarray}
imply the following recursion relations that are very useful:
\begin{eqnarray}
G_{k+1}(z) &=& G_{k-1}(z) + \frac{d+1+2k}{z} G_k(z),
\label{Eq:GkRecursion1}\\
G_k'(z) &=& 
-G_{k+1}(z) + \frac{k}{z} G_k(z) + G_1(z) G_k(z)
 = -G_{k-1}(z) - \frac{d+1+k}{z} G_k(z) + G_1(z) G_k(z),
\label{Eq:GkRecursion2}
\end{eqnarray}
for $k=0,1,2,3,\ldots$. In particular, one finds
\begin{eqnarray}
    \label{Eq:G1G-1}
    G_1(z) &=& G_{-1}(z) + \frac{d+1}{z}, \\
    \label{Eq:G2}
    G_2(z) &=& 1 + \frac{d+3}{z} G_1(z), \\
    \label{Eq:G3}
    G_3(z) &=& G_1(z) + \frac{d+5}{z} + \frac{(d+3)(d+5)}{z^2} G_1(z), \\
    \label{Eq:G1Prime}
    G_1'(z) &=& -1 - \frac{d+2}{z} G_1(z) + G_1(z)^2,
\end{eqnarray}
where a prime denotes derivative. 

In terms of the functions $G_k$ the generating function, enthalpy per particle, and the heat capacities per particle at constant pressure and volume defined in Eqs.~(\ref{Eq:Z0}), (\ref{Eq:EnthalpyPerParticle}), (\ref{Eq:cp}), and (\ref{Eq:cv}) can be written as
\begin{eqnarray}
Z &=& \frac{n}{m} G_{-1}(z), 
\label{Eq:Generatingf0}\\
h &=& m G_1(z), \\
c_\mathrm{p} &=& -k_B z^2 G_1'(z), \\
c_v &=& -k_B\left[ 1 + z^2 G_1'(z) \right],
\end{eqnarray}
respectively. For $d=3$, one has
\begin{equation}
G_1(z) = \frac{\textbf{K}_{3}(z)}{\textbf{K}_{2}(z)},
\end{equation}
which coincides with the function $G(\zeta)$ (with $\zeta=z$) defined on page~69 in~\cite{CercignaniKremer-Book}.

\section{Proofs of Lemmas~\ref{Lem:G-1},\ref{Lem:G1}, and \ref{Lem:hecpcv}}
\label{App:GProofs}

In this appendix we provide the proofs of Lemmas~\ref{Lem:G-1},\ref{Lem:G1}, and \ref{Lem:hecpcv} stated in Section~\ref{SubSec:Monotonicity}. For this, we recall the asymptotic behaviours of the functions $G_k$ and their first derivative in Eqs.~(\ref{Eq:GkNonrel}), (\ref{Eq:GkPrimeNonrel}), and (\ref{Eq:GkUltraRel}). Note that, together with Eq.~(\ref{Eq:G1Prime}), it also follows that $z^2 G_1'(z)\to -(d+1)$ for $z\to 0$. 

\proofof{Lemma~\ref{Lem:G-1}} We first use the asymptotic behaviours of $G_{-1}(z)$ which imply that $G_{-1}(z) \sim z/(d-1)$ for small $z\geq 0$ and $G_{-1}(z) = 1 - d/(2z) + d(d+2)/(8z^2) + {\cal O}(z^{-3})$ for large $z$; in particular $G_{-1}(0) = 0$ and $G_{-1}(z)\to 1$ as $z\to \infty$. It remains to show that $G_{-1}$ is strictly monotonously increasing. For this, we observe that $G_{-1}$ satisfies the differential equation
\begin{equation}
G_{-1}'(z) = -1 + \frac{d}{z} G_{-1}(z) + G_{-1}(z)^2,
\label{Eq:G-1prime}
\end{equation}
as can be deduced from Eqs.~(\ref{Eq:GkRecursion1}) and (\ref{Eq:GkRecursion2}). We know from the asymptotic behaviour that $G_{-1}'(z) > 0$ near $z=0$ and for large enough $z$. Assume $G_{-1}'$ has a zero at some $z^* > 0$. Then, it would follow that $G_{-1}''(z^*) = -d G_{-1}(z^*)/(z_*)^2 < 0$ which means $G_{-1}'$ can only cross zero from above, leading to a contradiction. Therefore, $G_{-1}'$ has no zeros and $G_{-1}$ is strict monotonously increasing, as claimed.
\qed

The proof of Lemma~\ref{Lem:G1} is very similar:

\proofof{Lemma~\ref{Lem:G1}} We know from the asymptotic behaviour that $G_k(z)\to \infty$ when $z\to 0$ and $G_k(z)\to 1$ when $z\to \infty$. Hence, it remains to prove the monotonicity property. We start with the case $k=1$. Using the asymptotic behaviour of $G_1'(z)$ we conclude that $G_1'(z)$ is negative for small or very large $z > 0$. Next, we claim that $G_1'(z)$ cannot have zeros on the interval $(0,\infty)$. Otherwise, if $z^*$ would be such a zero it would follow from Eq.~(\ref{Eq:G1Prime}) that $G_1''(z^*) = (d+2) G_1(z^*)/(z^*)^2 > 0$ which means $G_1'$ can only cross zero from below, leading to a contradiction.

Finally, by differentiating both sides of Eq.~(\ref{Eq:GkRecursion1}) with respect to $z$, it is easy to show that $G_{k-1}'(z)\leq 0$ and $G_k'(z) < 0$ imply that $G_{k+1}'(z) < 0$, so that $G_k'(z) < 0$ for all $k=1,2,3,\ldots$ follows by induction in $k$.
\qed

\proofof{Lemma~\ref{Lem:hecpcv}} Since $Z = n G_{-1}(z)/m$ and $h = m G_1(z)$ the statements for $Z$ and $h$ follow immediately from Lemmas~\ref{Lem:G-1} and~\ref{Lem:G1}, respectively. The statements for $h$ and $e$ follow from the definitions~(\ref{Eq:cp}) and (\ref{Eq:cv}) once the bounds~(\ref{Eq:cpcvBounds}) have been established, since these bounds imply that $c_\mathrm{p}$ and $c_v$ are positive. Clearly, the monotonicity and bound for $c_v$ follow from those of $c_\mathrm{p}$ since $c_v/k_B = c_\mathrm{p}/k_B - 1$.

Therefore, it remains to prove the bounds~(\ref{Eq:cpcvBounds}) for $c_\mathrm{p}/k_B$ and show that $c_\mathrm{p}$ is monotonous. For this, introduce the function
\begin{equation}
H(z):= -z^2 G_1'(z) = \frac{c_\mathrm{p}}{k_B}
\end{equation}
which is positive according to Lemma~\ref{Lem:G1}. Furthermore, we know from the asymptotic behaviour that $H(z)\to d+1$ and $H(z)\to d/2 + 1$ in the limits $z\to 0$ and $z\to \infty$, respectively. This establishes the claimed limits for $c_\mathrm{p}/k_B$ for $T\to 0$ and $T\to \infty$. Next, using Eq.~(\ref{Eq:G1G-1}) we can rewrite
\begin{equation}
H(z) = -z^2 G_{-1}'(z)  + d + 1 < d + 1,
\end{equation}
according to Lemma~\ref{Lem:G-1}, which proves the claimed upper bound for $c_\mathrm{p}/k_B$. To prove the lower bound, differentiate $H$ and use Eq.~(\ref{Eq:G1Prime}) to obtain the differential equation
\begin{equation}
H'(z) = -\frac{d}{z} H(z) 
 - 2G_1(z)\left[ \frac{d}{2} + 1  - H(z) \right].
\label{Eq:HDiffEq}
\end{equation}
Suppose $H(z)\leq d/2 + 1$ for some $z > 0$. Then, $H$ must have a minimum at some $z_* > 0$ where $H(z_*)\leq d/2 + 1$ and $H'(z_*) = 0$. However, this leads to a contradiction with Eq.~(\ref{Eq:HDiffEq}) since the left-hand side would be zero but the right-hand side negative at $z = z_*$. Therefore, $d/2 + 1 < H(z) < d + 1$ holds for all $z > 0$.

Finally, we prove that $H$ cannot have critical points. Suppose by contradiction that $z^* > 0$ is such that $H'(z^*) = 0$, then it would follow from Eq.~(\ref{Eq:HDiffEq}) and its first derivative that
\begin{equation}
H''(z^*) = \frac{d}{(z^*)^2} H(z^*) 
 - 2G_1'(z^*)\left[ \frac{d}{2} + 1 - H(z^*) \right]
 = \frac{d}{(z^*)^2}\frac{H(z^*)}{G_1(z^*)}\left[ G_1(z^*) + z^* G_1'(z^*) \right].
\end{equation}
Using the identity~(\ref{Eq:G1G-1}) one can check that
\begin{equation}
G_1(z) + zG_1'(z) = G_{-1}(z) + z G_{-1}'(z) > 0,
\end{equation}
which implies that any critical point of $H$ must be a strict local minimum of $H$. However, because $\lim\limits_{z\to 0} H(z) = d/2 + 1$ and $\lim\limits_{z\to\infty} H(z) = d + 1$ and the already established bounds on $H$, this is not possible.
\qed

\section{Proof of Lemma~\ref{Lem:nu}}
\label{App:ProofLemnu}

In this appendix we provide a proof of Lemma~\ref{Lem:nu} which generalizes the corresponding three-dimensional result in~\cite{mDmE88} to arbitrary dimension $d\geq 2$. In order to do so, we work in an orthonormal frame at $x$ such that $u = (1,0)$ and parametrize $p_1$ and $p_2$ by means of two hyperbolic angles $\chi_1,\chi_2\geq 0$ and two unit vectors $n_1,n_2\in S^{d-1}$, such that
\begin{equation}
p_a = m(\cosh\chi_a,\sinh\chi_a n_a),\qquad
a = 1,2.
\end{equation}
Then, one finds
\begin{equation}
\gamma = -\frac{u_\mu p_1^\mu}{m} = \cosh\chi_1,\qquad
d\mu_x(p_2) = f^{(0)}(x,p_2)\dvol_x(p_2) = m^{d-1}\alpha(x) (\sinh\chi_2)^{d-1} e^{-z\cosh(\chi_2)} d\chi_2 d\Omega(n_2),
\end{equation}
where $z := m/(k_B T(x)) > 0$. Furthermore, the relative speed $\mathrm{g} = |p_2-p_1|/m$ is given by
\begin{equation}
\mathrm{g} = \sqrt{2}\sqrt{\cosh\chi_1\cosh\chi_2 - 1 -\sinh\chi_1\sinh\chi_2 n_1\cdot n_2},
\end{equation}
from which one obtains the estimates
\begin{equation}
2\left( \cosh\chi_1\cosh\chi_2 - 1 \right) \leq \mathrm{g}^2 \leq 4\cosh\chi_1\cosh\chi_2.
\end{equation}
Here, the lower bound is restricted to the set $n_1\cdot n_2\leq 0$, while the upper one holds in general since $|n_1\cdot n_2|\leq 1$ and $\sinh\chi_a\leq \cosh\chi_a$. Using Eq.~(\ref{Eq:GradCutOff3}) and $\cosh\chi_1\geq 1$ one finds
\begin{equation}
c_1 (2\cosh\chi_1)^{\frac{\rho}{2}+1} (\cosh\chi_2 - 1)^{\frac{\rho}{2}+1} 
 1_{\{ n_1\cdot n_2\leq 0 \} }
\leq
k_1(p_1,p_2) = \mathcal{F}\sigma_T(\mathrm{g}) \leq c_2(\cosh\chi_1)^{\frac{\rho}{2}+1}\left( 1 + 2^{\rho+2}\cosh^{\frac{\rho}{2}+1}\chi_2 \right),
\label{Eq:k1Bound}
\end{equation}
where $1_{\{ n_1\cdot n_2\leq 0 \} }$ denotes the characteristic function associated with the set $n_1\cdot n_2\leq 0$, defined as being one on this set and zero otherwise. Integrating both sides over $p_2\in P_x^+(m)$ and taking into account Eq.~(\ref{Eq:Defnux}) yields
\begin{equation}
C_1(\cosh\chi_1)^{\frac{\rho}{2}+1} \leq \nu_x(p_1) \leq C_2(\cosh\chi_1)^{\frac{\rho}{2}+1},
\end{equation}
with
\begin{eqnarray}
C_1 &:=& 2^{\frac{\rho}{2}} c_1\mbox{Vol}(S^{d-1}) m^{d-1}\alpha(x)\int\limits_0^\infty (\cosh\chi_2 - 1)^{\frac{\rho}{2}+1}(\sinh\chi_2)^{d-1} e^{-z\cosh(\chi_2)} d\chi_2,\\
C_2 &:=& c_2\mbox{Vol}(S^{d-1}) m^{d-1}\alpha(x)\int\limits_0^\infty \left( 1 + 2^{\rho+2}\cosh^{\frac{\rho}{2}+1}\chi_2 \right)(\sinh\chi_2)^{d-1} e^{-z\cosh(\chi_2)} d\chi_2.
\end{eqnarray}
The continuity of $\nu_x$ follows from the continuity of $k_1$ in its first argument, the upper bound in Eq.~(\ref{Eq:k1Bound}) and Lebesgue's dominated convergence theorem. This concludes the proof of Lemma~\ref{Lem:nu}. Finally, we observe that the upper bound in Eq.~(\ref{Eq:k1Bound}) immediately yields the inequality~(\ref{Eq:HilbertSchmidt}).

\section{Orthogonal projection of  $p^\mu p^\nu$}
\label{App:Projection}

In this Appendix we compute the orthogonal projection of $p^\mu p^\nu$ onto $(\ker\mathcal{L})^\perp$. This generalizes the corresponding result obtained in \cite{JNET24} to arbitrary dimensions. We start with the following ansatz:
\begin{equation}
(p^\mu p^\nu)^\perp = p^\mu p^\nu + 2m b_1 u^{(\mu} p^{\nu)}
 + m\left( mb_2 + b_3 u^\alpha p_\alpha \right) u^\mu u^\nu
 + m\left( mb_4 + b_5 u^\alpha p_\alpha \right)\Delta^{\mu\nu},
\label{Eq:pportho}
\end{equation}
and choose the coefficients $b_j$ such that
\begin{eqnarray}
0 &=& \int\limits_{P_x^+(m)} (p^\mu p^\nu)^\perp f^{(0)}(x,p)
\dvol_x(p)
\nonumber\\
 &=& T^{\mu\nu} + 2m b_1 u^{(\mu} J^{\nu)}
  + m(mb_2 Z + b_3 u^\alpha J_\alpha) u^\mu u^\nu
  + m(mb_4 Z + b_5 u^\alpha J_\alpha) \Delta^{\mu\nu},
\end{eqnarray}
and
\begin{eqnarray}
0 &=& \int\limits_{P_x^+(m)} p^\alpha (p^\mu p^\nu)^\perp f^{(0)}(x,p)
\dvol_x(p)
\nonumber\\
 &=& T^{\alpha\mu\nu} + 2m b_1 u^{(\mu} T^{\nu)\alpha}
  + m(mb_2 J^\alpha + b_3 u_\beta T^{\alpha\beta}) u^\mu u^\nu
  + m(mb_4 J^\alpha + b_5 u_\beta T^{\alpha\beta}) \Delta^{\mu\nu}.
\end{eqnarray}
Using the expressions~(\ref{Eq:Generatingf0}), (\ref{Eq:T1Equilibrium}, (\ref{Eq:T2Equilibrium}), and (\ref{Eq:T3Equilibrium}) for $Z$, $J^\mu$, $T^{\mu\nu}$, and $T^{\alpha\mu\nu}$ one finds the conditions $b_1 = -G_1(z)$ and
\begin{equation}
M(z)\left( \begin{array}{c} b_2 \\ b_3 \end{array} \right) = \left( \begin{array}{c} G_1(z) + \frac{1}{z} \\ G_2(z) - \frac{1}{z}G_1(z) - 2G_1(z)^2 \end{array} \right),
\qquad
M(z)\left( \begin{array}{c} b_4 \\ b_5 \end{array} \right) = \frac{1}{z}\left( \begin{array}{c} -1 \\ G_1(z) \end{array} \right),
\label{Eq:bLinProblem}
\end{equation}
with the matrix $M(z)$ being
\begin{equation}
M(z) := \left( \begin{array}{cc}
 G_{-1}(z) & -1 \\
 -1 & G_1(z) - \frac{1}{z}
\end{array} \right).
\end{equation}
Using Eq.~(\ref{Eq:G-1prime}) and the result from Lemma~\ref{Lem:G-1} one has for the determinant of $M(z)$:
\begin{equation}
G_{-1}(z)\left( G_1(z) - \frac{1}{z} \right) - 1 = G_{-1}(z)^2 + \frac{d}{z} G_{-1}(z) - 1 = G_{-1}'(z) > 0,
\end{equation}
such that $M(z)$ is invertible. Solving Eq.~(\ref{Eq:bLinProblem}) yields the following results for the coefficients $b_i$ in Eq.~(\ref{Eq:pportho}):
\begin{eqnarray}
b_1 &=& -G_1(z),\\
b_2 &=& \frac{1}{z^2 G_{-1}'(z)}\frac{c_v}{k_B},\\
b_3 &=& \frac{1}{z^2 G_{-1}'(z)}\left[ (d+2)G_1(z)\frac{c_v}{k_B} - d\left( G_1(z) + \frac{1}{z} \right)\frac{c_\mathrm{p}}{k_B} \right],\\
b_4 &=& \frac{1}{z^2 G_{-1}'(z)},\\
b_5 &=& \frac{z G_1'(z) + G_1(z)}{z^2 G_{-1}'(z)}.
\end{eqnarray}
In particular, it follows from Eq.~(\ref{Eq:pportho}) that
\begin{eqnarray}
p^\alpha p^\beta\Delta_{\alpha\beta}{}^{\mu\nu} &=& (p^\alpha p^\beta)^\perp\Delta_{\alpha\beta}{}^{\mu\nu} ,
\\
p^\mu p^\nu u_\mu\Delta_\nu{}^\alpha &=&
(p^\mu p^\nu)^\perp u_\mu\Delta_\nu{}^\alpha - m G_1(z) p^\nu\Delta_\nu{}^\alpha,
\\
p^\mu p^\nu u_\mu u_\nu &=& (p^\mu p^\nu)^\perp u_\mu u_\nu + m c_1 p^\nu u_\nu + m^2 c_2, 
\end{eqnarray}
where we have introduced the projection onto the spatial trace-free part
\begin{equation}
\Delta_{\alpha\beta}{}^{\mu\nu} := \Delta_\alpha{}^\mu\Delta_\beta{}^\nu - \frac{1}{d}\Delta_{\alpha\beta}\Delta^{\mu\nu},
\end{equation}
and the coefficients $c_1$ and $c_2$ are given by
\begin{eqnarray}
c_1 &:=& 2b_1 - b_3 
 = \frac{d}{z^2 G_{-1}'(z)}\left[ G_1(z)\frac{c_v}{k_B} - \left( G_1(z) - \frac{1}{z} \right)\frac{c_\mathrm{p}}{k_B} \right],
\\
c_2 &:=& -b_2 = -\frac{1}{z^2 G_{-1}'(z)}\frac{c_v}{k_B}.
\end{eqnarray}

\section{General moment functionals}
\label{App:GeneralMomentIntegrals}

In this appendix, we define the functionals $\mathcal{W}_{n,k}[\phi]$ used in Section~\ref{Sec:Invariant}. For this, consider an arbitrary function $\phi$ of the variable $\gamma$ defined by
\begin{equation}
\gamma := -\frac{u^\mu p_\mu}{m},
\end{equation}
as in Eq.~(\ref{Eq:nuEstimate}). Then, one has the following expressions for its moments:
\begin{eqnarray}
\int\limits_{P_x^+(m)} \phi(\gamma) \dvol_x(p)
 &=& \mathcal{W}_{0,0}[\phi],
\label{W1}\\
\int\limits_{P_x^+(m)} p_\mu \phi(\gamma) \dvol_x(p)
 &=& m u_\mu\mathcal{W}_{1,0}[\phi],
\label{Wp}\\
\int\limits_{P_x^+(m)} p_\mu p_\nu\phi(\gamma) \dvol_x(p)
 &=& m^2\left( u_\mu u_\nu \mathcal{W}_{2,0}[\phi] + \frac{1}{d}\Delta_{\mu\nu} \mathcal{W}_{0,1}[\phi]\right),\label{Wpp}\\
\int\limits_{P_x^+(m)} p_\mu p_\nu p_\sigma\phi(\gamma) \dvol_x(p)
 &=& m^3\left( u_\mu u_\nu u_\sigma\mathcal{W}_{3,0}[\phi] + \frac{3}{d}\Delta_{(\mu\nu} u_{\sigma)} \mathcal{W}_{1,1}[\phi] \right),
 \label{Wppp}\\
\int\limits_{P_x^+(m)} p_\mu p_\nu p_\sigma p_\rho\phi(\gamma) \dvol_x(p)
 &=& m^4\left( u_\mu u_\nu u_\sigma u_\rho\mathcal{W}_{4,0}[\phi] + \frac{6}{d}\Delta_{(\mu\nu} u_{\sigma} u_{\rho)} \mathcal{W}_{2,1}[\phi] + \frac{3}{d(d+2)}\Delta_{(\mu\nu}\Delta_{\sigma\rho)} \mathcal{W}_{0,2}[\phi] \right),
\label{Wpppp}
\end{eqnarray}
where the functions $\mathcal{W}_{n,k}[\phi]$ are defined as
\begin{eqnarray}
\mathcal{W}_{n,k}[\phi]
&:=& m^{d-1}\mbox{Vol}(S^{d-1}) \int\limits_0^\infty \phi(\cosh\chi)\cosh^n\chi \sinh^{2k+d-1}\chi d\chi
\nonumber\\
 &=& m^{d-1}\frac{2\pi^{d/2}}{\Gamma\left( \frac{d}{2} \right)} \int\limits_1^\infty \phi(\gamma)\gamma^n\left( \gamma^2 - 1 \right)^{k-1 + \frac{d}{2}} d\gamma.
\label{Eq:Wnkphi}
\end{eqnarray}
For $d=3$ and $\phi(\gamma) = e^{-z\gamma}$, these integrals reduce to the corresponding moment integrals $\mathcal{W}_{n,k}$ introduced in Ref.~\cite{aGaM2023}.

\bibliographystyle{unsrt}
\bibliography{refs_kinetic}

\end{document}